\documentstyle[12pt]{article}
\newcommand{\beq}{\begin{equation}}
\newcommand{\eeq}{\end{equation}}

\def\quaft{{\textstyle {{{1}\over{4\pi\alpha'}}} }}
\def\half{{\textstyle{1\over2}}}

\def\lh{{{il}\over{2}}}

\def\p1half{{\textstyle{{{p+1}\over{2}}}}}

\def\23phalf{{\textstyle{{{23-p}\over{2}}}}}

\begin{document}
\pagestyle{empty}
\begin{titlepage}

\bigskip
\hskip 3.7in{\vbox{\baselineskip12pt
\hbox{PSU-TH-228}\hbox{hep-th/0006014}}}

\bigskip\bigskip\bigskip\bigskip
\centerline{\large\bf Effective String Tension and Renormalizability:}
\centerline{\large\bf String Theory in a Noncommutative Space}
\bigskip\bigskip
\bigskip\bigskip
\centerline{\bf Shyamoli Chaudhuri and Eric Novak
\footnote{shyamoli@phys.psu.edu,novak@phys.psu.edu}
}
\centerline{Physics Department}
\centerline{Penn State University}
\centerline{University Park, PA 16802}
\date{\today}

\bigskip\bigskip
\begin{abstract}
\noindent
We show that the one loop amplitudes of open and closed string theory 
in a constant background two-form tensor field are characterized by an 
effective string tension larger than the fundamental string tension, and 
by the appearance of antisymmetric and symmetric noncommutativity parameters.  
We derive the form of the phase functions normalizing planar and nonplanar 
tachyon scattering amplitudes in this background, verifying the decoupling
of the closed string sector in the regime of infinite momentum transfer. 
We show that the functional dependence of the phase functions on the 
antisymmetric star product of external momenta permits interpretation 
as a finite wavefunction renormalization of vertex 
operators in the open string sector. 
Using world-sheet duality we clarify the regimes of finite and zero 
momentum transfer between boundaries, demonstrating the existence of poles 
in the nonplanar amplitude when the momentum transfer equals the mass of 
an on-shell closed string state. Neither noncommutativity parameter has any 
impact on the renormalizability of open and closed string theory in the 
Wilsonian sense. We comment on the relationship to noncommutative scalar 
field theory and the UV-IR correspondence. 

\end{abstract} \end{titlepage} 
\pagestyle{plain}

\section{Introduction}
\label{sec:intro}

The idea that the coordinates of spacetime do not commute
at sufficiently small distance scales has received new scrutiny
with the discovery of nonperturbative backgrounds
of String/M theory that are noncommutative spacetimes
\cite{witten,cds}. The worldvolume of a Dbrane with constant 
background two-form tensor gauge fields is a simple and 
concrete example of a noncommutative spacetime in which live
gauge and matter fields \cite{callan,ft,ch,jabbari,schomerus,sw}.
Motivated in part by the puzzling mix of ultraviolet and infrared 
effects recently observed in noncommutative scalar field theories 
\cite{srm}, we will revisit the one-loop amplitudes of open and 
closed bosonic string theory in the presence of constant two-form
background fields \cite{ft,callan}.\footnote{Since this project was
begun, many papers have appeared on this subject with overlapping
results \cite{dorn,sang,bilal,gomis,liumic}. We will point out the 
differences at appropriate places in the text and in the conclusions.}
We will attempt to \lq\lq discover" possibly unusual properties of 
quantum field theory in this noncommutative space by deriving the 
field theory limit, in both open and closed string channels, of the 
one-loop tachyon scattering amplitude of bosonic string theory in 
a constant background field. The result is a surprise in many 
respects.

In section II, we point out that the \lq\lq open string metric" 
\cite{sw} appearing in the $B$ dependent normalization of the one-loop
vacuum amplitude, originally computed in \cite{ft,callan}, has a 
natural world-sheet interpretation as the effective string tension
of open string theory in a constant background two-form
tensor field. The effective string tension provides the natural scale 
with respect to which we measure momenta and energies in this theory. 
We will find that the effective string tension is always {\em larger}
than the bare fundamental string tension, for nonvanishing background 
two-form fields. The scattering amplitudes of the theory are further 
characterized by the presence of antisymmetric and symmetric noncommutativity 
parameters which will be explored in sections IV and V. The 
noncommutativity scale is a priori distinct from the effective string
tension and is associated with wavefunction renormalization for states 
in the open string sector.\footnote{A different notion of an effective
string scale is described in \cite{kawaiet}, where the effective scale of
string-like excitations in the IIB matrix model on a Von Neumann lattice 
is found to be identical to the noncommutativity scale of NCYM theory.} 
We will find in section III that the noncommutativity parameters only 
enter the {\em finite} part of the Greens function on the annulus, thereby 
determining the vertex operator algebra and the external momentum 
dependent normalizations of the amplitudes, but with no bearing on 
the usual renormalizability 
properties of open and closed string theory in this background. The
effective string tension, on the other hand, does appear in the short
distance divergence of the Greens function for closely separated
sources on the boundary of the world-sheet. 
This log divergence is absorbed in a renormalization 
of the bare open string coupling, precisely analogous to the case of free 
strings \cite{poltorus,polbook}.

The main distinction between our string theory interpretation and that of 
noncommutative scalar field theory is due to the application of open-closed 
string world-sheet duality. An analysis of the nonplanar amplitude in the open
string channel indeed displays both the momentum dependent phase functions and 
the dependence on internal momentum transfer between boundaries, found in the 
corresponding nonplanar graphs of a noncommutative scalar field theory 
\cite{srm}. 
Momentum transfer between boundaries in string theory is addressed by using
world-sheet duality to express the nonplanar amplitudes in the closed string
channel. The potentially puzzling regime of zero momentum transfer 
is dominated by the exchange of a zero momentum massless
state in the closed string sector. We will show that this limit is benign. 
The only renormalization necessary in our theory is the usual renormalization
of the open string coupling, and the ${\bf p}$$\to$$0$ limit of zero momentum 
transfer therefore commutes with taking the ultraviolet cut-off on momenta 
in the open string channel to infinity. It should be emphasized that in 
using world-sheet duality to rewrite the string amplitude in terms of the 
massless field theory limit of the closed string sector, we are outside the 
domain of the original noncommutative field theory which corresponds to 
the massless limit of the open string sector. Our conclusion is that open 
and closed string theory in a background two-form tensor field, albeit in 
a noncommutative space, displays ordinary 
Wilsonian behavior as regards renormalizability. The finite wavefunction 
renormalization of external states and the finite renormalization of the 
string tension are the main remnant signals of the noncommutative nature 
of the embedding spacetime.\footnote{We are careful in distinguishing
{\em finite} renormalization effects, as in the generation of an effective 
string tension and distinct noncommutativity scale from the bare parameters,
$\alpha^{\prime}$ and $B^{\mu\nu}$, in the world-sheet action, from what 
we refer to as 
{\em Wilsonian} renormalization: the rescaling of the infinite bare
parameters of a theory to their finite physical values due to quantization 
of the ultraviolet, or short distance, degrees of freedom. The renormalization 
of the open string coupling constant--- and of the 
world-sheet cosmological constant, which is renormalized to zero--- are 
Wilsonian renormalizations, analogous to Wilsonian renormalization in a 
quantum field theory. Note, also, the clarifications in footnote 12.}

In what follows, we give a path integral derivation of both the one-loop vacuum 
amplitude and the planar, and nonplanar, one-loop tachyon scattering amplitudes 
in open and closed bosonic string theory in a constant background
two-form tensor field. A path integral derivation of the $N$ point closed string
tachyon scattering amplitude on the torus for free strings was given in \cite{poltorus}. 
In sections II and III, we adapt this calculation to open and closed string 
theory in a constant background two-form tensor field. The method enables us
to derive the Greens function and the renormalized scattering amplitudes inclusive 
of normalization, and of all of the background field dependence. 
In order to make the paper
self-contained, we include in the appendix a pedagogical discussion of the method
in \cite{poltorus} with necessary extensions used by us to calculate the bulk Greens 
function on the annulus in constant background fields.

In section IV we derive the precise form of the momentum dependence of the phase 
functions normalizing the planar one-loop amplitudes, assuming the simplifying kinematics 
of forward momentum transfer. While these are indeed functions of the star products of
the external momenta, the phase functions on the annulus are found to be more complex than 
the simple exponential found in noncommutative scalar field theory \cite{flk,srm}, and 
in the vertex operator algebra of 
open string theory on the disk \cite{sw}. 
The antisymmetric noncommutativity parameter
is interpreted as giving a wavefunction renormalization for states in the open string 
sector. Finally, in section V, we derive the nonplanar amplitude verifying the presence 
of a phase proportional to the momentum transfer between boundaries. Using world-sheet 
duality, we express the string amplitude in terms of closed string variables demonstrating 
the existence of a pole in the amplitude when the momentum transfer equals 
the mass of 
an on-shell closed string state. We summarize our results in the
conclusions, making a comparison with the results of other authors
in recent papers overlapping our work.

\section{Effective String Tension and Vacuum Amplitude}
\label{sec:vacu}

The one-loop vacuum amplitude in the presence of constant $B$ field
has been derived by previous authors using both the path integral 
method \cite{ft} and the background field technique \cite{callan}. 
In this section,
we obtain this result following the treatment of the one-loop torus
path integral for free strings given in \cite{poltorus}, using the 
same technique to evaluate the Greens function and one-loop scattering 
amplitudes in following sections of the paper. Thus, we adapt the free 
closed string calculation in \cite{poltorus} to open and closed string 
theory in constant 
background fields in a flat spacetime. We will assume a spacetime metric
of Euclidean signature, commenting briefly on the necessary modifications 
of this analysis in the case of a Lorentzian metric. 
We point out that the \lq\lq open string metric" \cite{sw}, 
which appears in the $B$ dependent normalization of the one-loop vacuum 
amplitude, has a natural world-sheet interpretation as the effective 
string tension of open and closed string theory in the presence of a 
background $B$ field. Masses and couplings in the low energy theory are 
measured in units of the effective tension, 
$(\alpha^{\prime}_{{\rm eff.}})^{-1/2}$,
as opposed to the bare string tension, $(\alpha^{\prime})^{-1/2}$, with 
$(\alpha^{\prime}_{{\rm eff.}})^{-1/2}$ $>$ $(\alpha^{\prime})^{-1/2}$
for non-vanishing $B$ field. It should be emphasized that the effective
string tension is a distinct energy scale from the noncommutativity 
scale, which enters in the vertex operator algebra and scattering
amplitudes of the theory.
These are discussed in later sections. The effective string scale in a 
background $B$ field may also be interpreted fruitfully as a finite 
renormalization of the fundamental string tension.

We begin with the world-sheet action:
\begin{equation} 
S  =  {{1}\over{4\pi\alpha^{\prime}}} \int_M d^2 \sigma ( {\sqrt{g}} 
   g^{ab} g_{\mu\nu}
+ i \epsilon^{ab} B_{\mu\nu} ) \partial_a X^{\mu} \partial_b X^{\nu}
+ i \int_{\partial M} d\lambda ( A_{\mu}
{{d}\over{d\lambda}} X^{\mu} ) 
\quad ,
\label{eq:action}
\end{equation}
where $g_{\mu\nu}$$=$$\delta_{\mu\nu}$, $A_{\mu}$ is a constant 
background vector field on the branevolume, and 
$B_{\mu\nu}$ is a real, antisymmetric, and constant, background
tensor field.\footnote{Note that $g$ need not be a flat
spacetime metric, the derivation is identical for arbitrary
constant $g$. In writing Eq.\ (\ref{eq:action}), the constant
background fields $g_{\mu\nu}$, $B_{\mu\nu}$, and $A_{\mu}$, have 
been assumed dimensionless, while the embedding coordinates
and external momenta are dimensionful.}
The $B$ field is restricted to the worldvolume of 
a Dpbrane with $p$ assumed to be an odd integer. Our results 
could therefore be extended to the supersymmetric type I string 
theory 
with its spectrum of odd $p$ Dpbranes. Note, however, that 
unlike the type I case, the eigenvalues of the constant $B$ field 
in the oriented bosonic string theory are not necessarily quantized 
and can take arbitrary values.

Let us label the directions parallel to the branevolume, 
$\mu$$=$$0$, $\cdots$,
$p$. The $25$$-$$p$ directions transverse to the brane volume 
satisfy Dirichlet boundary conditions. We will assume that the 
$B$ field has been brought to block diagonal form by a spacetime 
transformation on the background fields \cite{callan}, with 
next-to-diagonal 
entries $b_{(m)}$$=$$B_{2m,2m+1}$, $m$$=$$0$, $\cdots$, $(p-1)/2$. 
Note that the background preserves the Weyl invariance of the
action. Thus, gauge fixing to conformal gauge gives the same
contribution to the path integral from reparameterization ghost fields
as in the quantization of free strings. 
It can be shown that the $p$$+$$1$ dimensional
worldvolume of a Dpbrane is an even dimensional 
noncommutative space, whose coordinates satisfy the 
commutation relation \cite{ch}:   
\begin{equation}
[X^{\mu},X^{\nu}] = 2\pi i \alpha^{\prime} ({\cal M}^{-1} {\cal F} )^{\mu\nu}
\equiv i \Theta^{\mu\nu} \quad .
\label{eq:comm}
\end{equation}
The basic observation of a nonvanishing commutator for the zero 
modes was originally made in \cite{callan}.
For clarity, we consider a single Dpbrane and set the Maxwell 
term in the gauge invariant
combination of two-form background fields to zero. Thus,
${\cal F}$$\equiv$$B$$-$$2\pi\alpha^{\prime}F$$=$$B$, and the matrix,
${\cal M}$, takes block-diagonal form: 
${\cal M}_{\mu\nu} $$=$$ \delta_{\mu\nu} $$-$$ 
B_{\mu}^{\lambda} B_{\lambda\nu} $. The indices $\mu$, $\nu$, 
run from $0$, $\cdots$, $p$. In section 5 and the conclusions, 
we return to this issue, explaining briefly the modifications 
to our analysis for the unoriented bosonic string theory with 
$2^{13}$ D25branes and, potentially, a constant background for
the Yang-Mills field strength.

The commutation relations given above follow from the
Poisson brackets of the $X_{\mu}$ with their 
canonically conjugate momenta, $P_{\mu}$, imposing as a constraint
the boundary conditions:
\begin{equation}
g_{\mu\nu} {\sqrt{g}} n^a \partial_a X^{\nu} +
i B_{\mu\nu} 
t^a \partial_a X^{\nu} = 0 \quad , 
\label{eq:bndry}
\end{equation}
where $n^a$ and $t^a$ are, respectively, unit 
normal and tangent vectors to the boundary.
Thus, the theory can, at least in principle, 
be quantized exactly in a canonical framework.
In this paper, we will investigate the amplitudes of open and 
closed string theory in a constant background $B$ field 
by direct Lagrangian path integral evaluation.
We will find that the path integral computation
enables a derivation of the one-loop scattering amplitudes
with the normalization--- and all of the background 
field dependence--- explicit. The results obtained are on the same
precise footing as the one-loop amplitudes for free strings.

The gauge fixing of the Polyakov path integral is well-known 
and has been described elsewhere \cite{poltorus}.
We begin with the conformal gauge fixed 
expression for the annulus in constant background $B$ field
with both boundaries on the same Dpbrane.
It is convenient to use a complex basis, $Z_m$$=$$X^{2m}$$+
$$iX^{2m+1}$, $m$$=$$0$, $\cdots$, $(p-1)/2$, and its complex conjugate
coordinate, ${\bar Z}_m$, with $b_{(m)}$$\equiv$$B_{2m,2m+1}$.
Then the commutation relations take the simple form:
\begin{equation}
[Z_m, {\bar Z}_n] = {{4 \pi \alpha^{\prime} 
  b_{(m)}}\over{1+b_{(m)}^2 }} \delta_{mn} \equiv 2 \theta_{(m)} \delta_{mn} 
 \quad ,
\label{eq:noncom}
\end{equation}
and the boundary conditions on the complex scalars are given by
\begin{eqnarray}
l \partial_2 Z_m =&& - b_{(m)} \partial_1 Z_m 
\nonumber\\ 
l \partial_2 {\bar Z}_m =&&  b_{(m)} \partial_1 {\bar Z}_m 
\quad .
\label{eq:motion}
\end{eqnarray}
The amplitude takes the form:
\begin{equation}
 {\cal A} = \int_0^{\infty}  \left [ {{dl}\over{l}} \eta^2(\lh) \right ] 
   [\eta (\lh )]^{p-25} 
     \prod_{m=0}^{(p-1)/2} \int [d\delta Z_m]
[d \delta {\bar Z}_m] e^{- 
 \quaft \int d^2 \sigma ( {\sqrt{g}} 
   g^{ab} - \epsilon^{ab} b_{(m)} ) 
 \partial_a Z_m \partial_b {\bar Z}_m - \mu_0 \int d^2 \sigma {\sqrt{g}} }
 \quad ,
\label{eq:annulus}
\end{equation}
where $\mu_0$ is the bare world-sheet cosmological constant.
The measure in the functional integral has been evaluated
with respect to the fiducial cylinder metric,
$ds^2$$=$$l^2(d \sigma^1)^2$$+$$(d\sigma^2)^2$,
where $l$ is the length of the boundary and 
$0$$\le$$\sigma^a$$\le$$1$, $a$$=$$1$, $2$. 
The first factor in square brackets is the remnant of the 
functional integration over metrics, obtained upon gauge fixing 
world-sheet reparameterizations \cite{poltorus}. The second factor 
arises from the functional
integration over $25$$-$$p$ Dirichlet embedding coordinates. 
It remains to compute the contribution from the complex scalars 
satisfying the boundary conditions given in Eq.\ (\ref{eq:bndry}).

The Laplacian on scalars on the annulus takes the form 
$\Delta$$=$$l^{-2}\partial_1^2 $$+$$ \partial_2^2$.
We expand in a basis of eigenfunctions of the Laplacian satisfying
the given boundary conditions: 
\begin{eqnarray}
Z =&& z_0 + \sum_{n_1=-\infty}^{\infty} \sum_{n_2=1}^{\infty} z_{n_1n_2} \Psi_{n_1n_2} +
 \sum_{n_1=-\infty}^{\infty} z_{n_10} \Psi_{n_10} 
\nonumber\\
{\bar Z} =&& {\bar z}_0 + \sum_{n_1=-\infty}^{\infty} \sum_{n_2=1}^{\infty} {\bar z}_{n_1n_2} 
{\bar\Psi}_{n_1n_2} +
 \sum_{n_1=-\infty}^{\infty} {\bar z}_{n_10} {\bar\Psi}_{n_10} 
\quad,
\label{eq:expa}
\end{eqnarray}
where the complex basis functions, $\Psi$, take the form:
\begin{eqnarray}
\Psi_{n_1n_2} &&= {{1}\over{{\sqrt{l}}}} 
 {\rm exp }(2\pi i n_1 \sigma^1) \left ( {\rm Cos} (n_2 \pi \sigma^2)
- {{i\alpha n_1}\over{n_2 }}{\rm Sin} (n_2 \pi \sigma^2 ) \right )\cr  
{\bar \Psi}_{n_1n_2} &&= {{1}\over{{\sqrt{l}}}} 
 {\rm exp} (2\pi i n_1 \sigma^1) \left ( {\rm Cos} (n_2 \pi \sigma^2)
+ {{i\alpha n_1}\over{n_2}} {\rm Sin} (n_2 \pi \sigma^2 ) \right ) \cr
\Psi_{n_10} &&= {{1}\over{{\sqrt{l}}}} 
 {\rm exp}[ 2\pi i n_1 (\sigma^1 - {{b}\over{l}} \sigma^2) ] \cr
{\bar \Psi}_{n_10} &&= {{1}\over{{\sqrt{l}}}} 
 {\rm exp} [2\pi i n_1 ( \sigma^1 + {{b}\over{l}} \sigma^2) ] 
\quad ,
\label{eq:set}
\end{eqnarray}
and the parameter $\alpha$$=$$2b/l$. The eigenfunctions
correspond, respectively, to eigenvalues:
\begin{eqnarray}
\omega_{n_1n_2} &&= {{4\pi^2}\over{l^2}} n_1^2 + \pi^2 n_2^2 \nonumber \\ 
\omega_{n_10} &&=  {{4\pi^2n_1^2}\over{l^2}}(1 + b^2 ) 
 \quad ,
\label{eq:determ}
\end{eqnarray}
where the subscripts take values in the 
range $-\infty$$\le$$n_1$$\le$$\infty$, 
$n_2$$\ge$$1$. In the limit of zero $B$ field, we recover the eigenfunctions
and eigenspectrum of a complex scalar satisfying Neumann boundary conditions. 
A natural choice of reparameterization invariant norm
on the space of complex eigenfunctions is given by the 
orthogonality relations:
\begin{eqnarray}
\int d^2 \sigma {\sqrt {g}} 
  {\bar\Psi}_{n_1^{\prime}n_2^{\prime}} \Psi_{n_1n_2} 
&&= C_{n_1n_2} \delta_{n_1^{\prime},-n_1} 
 \delta_{n_2^{\prime}n_2}  \nonumber\\
\int d^2 \sigma {\sqrt {g}} 
  {\bar\Psi}_{n_1^{\prime}0} \Psi_{n_10} 
&&= C_{n_10} \delta_{n_1^{\prime},-n_1}  
\nonumber\\
\int d^2 \sigma {\sqrt {g}} 
  {\bar\Psi}_{n_1^{\prime}n_2} \Psi_{n_10} 
&&= \int d^2 \sigma {\sqrt {g}} 
  {\bar\Psi}_{n_1^{\prime}0} \Psi_{n_1n_2} =0  \quad , 
\label{eq:norm}
\end{eqnarray}
where the orthogonality coefficients are given by:
\begin{equation}
C_{n_1n_2} = \half (1 - n_1^2\alpha^2/n_2^2) ,
\quad\quad \quad C_{n_10} = 
e^{-i \pi n_1 \alpha}
  {{{\rm Sin}(\pi n_1\alpha)}\over{\pi n_1 \alpha}}  \quad .
\label{eq:params}
\end{equation}
Note that the coefficients are not positive definite. This is the
first indication of the unusual properties of open string theory
in a constant background $B$ field in a Euclidean spacetime: 
we will find that while
the orthogonality coefficients drop out of the expression for 
the vacuum amplitude, they play a crucial role in the scattering
amplitude. 
It should be noted that in a spacetime of Lorentzian 
signature--- with true {\em spacetime} noncommutativity, the
orthogonality coefficients would be positive definite.
We return to this point in the conclusions.

The normalization of the vacuum amplitude will be 
obtained as in \cite{poltorus}. We use the reparameterization 
invariant norm for a free complex scalar:
\begin{eqnarray}
|\delta Z|^2 =&& \int d^2\sigma {\sqrt{g}} (\delta\bar{Z})(\delta Z) 
\nonumber\\
=&& \sum_{n_1=-\infty}^{\infty} \sum_{n_2=1}^{\infty}
C_{n_1n_2} \delta {\bar z}_{-n_1n_2} \delta z_{n_1n_2}  +
\sum_{n_1=-\infty}^{\infty} 
C_{n_10} \delta {\bar z}_{-n_10} \delta z_{n_10} 
\quad , 
\label{eq:normsac}
\end{eqnarray}
substituting the basis functions given in Eq.\ (\ref{eq:set}).
Following \cite{poltorus}, we assume that the measure in the tangent
space to the space of embeddings is ultralocal. Namely, the functional 
integral over embeddings, $Z(\sigma^a)$, is the 
product of ordinary integrals defined at some base point, $\sigma^a$, 
on the world-sheet, followed by an integration of the location of the 
base point in the domain $0$$\le$$\sigma^a$$\le$$1$. The measure for the
zero modes is obtained in the usual way setting:
\begin{equation}
\int d\delta Z d\delta {\bar Z} ~
e^{-|\delta Z|^2/2} \equiv 1 = \int [d z_0][d{\bar z}_0] 
\prod_{n_1,n_2} \int dz_{n_1n_2} d{\bar z}_{n_1n_2} dz_{n_10} 
d{\bar z}_{n_10} ~ J ~ e^{-|\delta Z|^2/2} 
\quad ,
\label{eq:ultra}
\end{equation}
where $J$ is the Jacobian from the change of variables in Eq.\
(\ref{eq:normsac}). Performing 
the Gaussian integrations explicitly determines this as: 
\begin{equation}
J = {{l}\over{2\pi}} ~
\prod_{n_2=0}^{\infty} 
\prod_{n_1=-\infty}^{\infty\prime} \left ( {{ C_{n_1n_2} }\over{2\pi}}\right )
\quad .
\label{eq:jacobian}
\end{equation}
Thus, the path integral for a single complex scalar in constant background
field takes the form:
\begin{eqnarray}
&&\int [d\delta Z] [d\delta {\bar Z}] e^{- \quaft \int d^2 \sigma {\sqrt {g}}
 ~ {\bar Z} \Delta Z } = 
\left ( {{l}\over{2\pi}} \right )
\prod_{n_1=-\infty}^{\infty \prime} 
\prod_{n_2=0}^{\infty} 
{{ C_{n_1n_2} }\over{2\pi}}
\nonumber\\
&& \quad\quad \quad \times \int d z_{n_1n_2} d{\bar z}_{n_1n_2} 
 e^{- \quaft \left ( \sum_{n_1=-\infty}^{\infty} \sum_{n_2=0}^{\infty}
   \omega_{n_1n_2} C_{n_1n_2} {\bar z}_{n_1n_2} z_{n_1n_2} \right ) }
\quad ,
\label{eq:funcint}
\end{eqnarray}
where the prime denotes exclusion of the $n_1$$=$$n_2$$=$$0$ mode 
from the infinite product. Performing the integrals as above, we
obtain the following expression for the amplitude:
\begin{equation}
 {\cal A} = i V_{p+1} 
   \int_0^{\infty} \left [ {{dl}\over{l}} \eta^2(\lh) \right ] 
\eta (\lh )^{-(25-p)} 
({{l}\over{2\pi}})^{(p+1)/2}
\left [ \prod_{m=0}^{(p-1)/2} 
\prod_{n_1=-\infty}^{\infty} \prod_{n_2=0}^{\infty} (2\pi)
{{ \omega^{(m)}_{n_1n_2}}\over{4\pi^2\alpha^{\prime}}} \right ]^{-1}  \quad .
\label{eq:rannulus}
\end{equation}
Note that the Jacobian from the change of variables is cancelled 
against a similar term arising from the Gaussian integration, such 
that the orthogonality coefficients are absent from the final 
expression. The functional determinant of the Laplacian is evaluated 
using zeta function regularization. The result is:
\begin{equation}
 {\cal A} = i V_{p+1} {\rm det} \left ( {\bf 1} + {\bf B} \right ) 
    \int_0^{\infty} {{dl}\over{l}} 
      (4\pi^2 \alpha^{\prime} l)^{-(p+1)/2} \eta (\lh )^{-24}  
 \quad ,
\label{eq:resannulus}
\end{equation}
in agreement with the references \cite{ft,callan}. 
The renormalized value of the world-sheet cosmological constant has been
set to zero as in \cite{poltorus}.
The sum over the $n_2$$>$$0$ modes gives $p$$+$$1$ powers of the $\eta$ 
function, analogous to the result for 
$p$$+$$1$ free scalars with Dirichlet boundary conditions. The $B$ dependent 
pre-factor 
in the amplitude arises from the $n_2$$=$$0$ modes as follows:
\begin{equation}
 \prod_{m=0}^{(p-1)/2} \prod_{n_1=-\infty}^{\infty \prime} 
\left [ {{ \omega^{(m)}_{n_10}}\over{2\pi\alpha^{\prime}}} \right ]^{-1} 
= \prod_{m=0}^{(p-1)/2} \prod_{n_1=1}^{\infty} 
\left [ {{ \omega^{(m)}_{n_10}}\over{2\pi\alpha^{\prime}}} \right ]^{-2} 
= (2\pi \alpha^{\prime} l^2 )^{-(p+1)/2}  
{\rm det} ({\bf 1} + {\bf B} ) 
\quad ,
\label{eq:deter}
\end{equation}
where the prime denotes exclusion of the $n_1$$=$$0$ mode from the infinite 
product, and we have used the identity:
\begin{equation}
{\rm ln} ~ {\rm det} \{ \cdots \} = - \lim_{s\to 0} {{d}\over{ds}} \left \{ 
[{{2\pi}\over{l^2 \alpha^{\prime}}}(1+b^2)]^{-s}
\sum_{n_1=1}^{\infty} n_1^{-2s} \right \} = \zeta (0) ~ {\rm ln} 
~ [{{2\pi}\over{l^2 \alpha^{\prime}}}(1+b^2)]
- 2 \zeta^{\prime} (0) = \half {\rm ln} [{{2\pi l^2 
      \alpha^{\prime}}\over{(1+b^2)}}]
\quad .
\label{eq:nfpr}
\end{equation}
Here, $\zeta(0)$ is the continuation of the ordinary Riemann zeta function 
to its value with zero argument. In the limit of zero $B$ field, we recover
the contribution from the $n_2$$=$$0$ modes of $p$$+$$1$ real scalars 
satisfying Neumann boundary conditions \cite{poltorus}.

It is interesting to interpret the $B$ dependent normalization of the 
one-loop vacuum amplitude given in Eq.\ (\ref{eq:resannulus}).
Notice that the only change from the one-loop vacuum amplitude for free 
strings is that the fundamental string tension, 
$\tau_F$$=$${{1}\over {2\pi\alpha^{\prime}}}$,  
has been rescaled to an {\em effective} string tension:
\begin{equation}
\tau_{{\rm eff.}}^{(p+1)/2} = 
(2 \pi \alpha^{\prime}_{{\rm eff.}})^{-(p+1)/2} \equiv 
    \left [ {\rm det}({\bf 1} + {\bf B}) (2\pi \alpha^{\prime})^{-(p+1)/2} 
\right ]
\quad .
\label{eq:efft}
\end{equation}
Thus, open string 
theory in a constant background $B$ field probes 
shorter distance scales or, equivalently, higher momentum scales, 
with an ultraviolet cutoff of $O({\sqrt{\alpha^{\prime}_{{\rm eff.}}}})$.
The coupling to constant background fields effectively {\em raises} 
the string scale.
We will see this feature repeated in the scattering amplitudes: the
fundamental string scale does not appear in the scattering amplitudes 
except in the particular 
combination given in Eq.\ (\ref{eq:efft}). In addition, new parameters 
associated with noncommutativity are present in the planar and
nonplanar scattering amplitudes. The same is true of the low 
energy spacetime actions for massless fields derived from the string 
theory. They are characterized by the appearance of two finitely 
distinct energy scales: $(\alpha^{\prime}_{{\rm eff.}})^{-1/2}$, 
and a noncommutativity scale, $\Theta^{-1/2}$, to be defined below. 
We will find that while 
$(\alpha^{\prime}_{{\rm eff.}})^{-1/2}$ is always above the fundamental
string scale, the noncommutativity scale can be lower or higher 
depending on the specific values of the background fields. We return to 
this point in the conclusions.

\section{Greens Function on the Annulus}
\label{sec:greens}

The main ingredient required for the computation of planar and 
nonplanar one-loop amplitudes is the bulk Greens function for 
two sources on the annulus. In this section, we perform this 
computation, examining also the divergences in the Greens function 
for closely separated sources on the boundary of the world-sheet. 
We will find that the requisite renormalization of the open string
coupling proceeds much the same way as for free string theory, 
except for the appearance of noncommutativity parameters
in the {\em finite} part of the 
Greens function. This will lead to momentum dependent phase 
functions normalizing the planar and nonplanar scattering 
amplitudes. These are derived in the following section.

Consider the annulus with $N$ open string tachyon vertex
operator insertions:
\begin{equation}
V(k_i) = g \int_{\partial M} d s {\sqrt{g}}
  e^{i ({\bf k} \cdot {\bf {\bar Z}} + {\bf{\bar k}} \cdot {\bf Z}) } \quad , 
\label{eq:tachy}
\end{equation}
corresponding to delta function sources,
$J^m(\sigma^a)$$=$$k^m_i \delta (\sigma^1 - \sigma^1_i)$,
${\bar J}^m(\sigma^a)$$=$${\bar k}^m_i \delta (\sigma^1 - \sigma^1_i)$,
inserted at locations 
$\sigma^1_i$, $i$$=$$1$, $\cdots$, $N$, on the boundary
of the world-sheet, with momentum, 
$\sum_{m=0}^{(p-1)/2} k^m_i{\bar k}^m_i$$=$${\bf p}_i^2/4$. 
The factor of $4$ comes from our use of a complex basis for momentum 
vectors defined below. Here $g$ is the open string coupling. The 
contribution from a single complex scalar to the path integral over 
world-sheets coupled to external sources takes the form \cite{polbook}:
\begin{eqnarray}
&&\int [d\delta Z] [d\delta {\bar Z}] e^{\quaft \int d^2 \sigma {\sqrt {g}}
 ~ {\bar Z} \Delta Z + i \int d^2 \sigma {\sqrt{g}}
  ({\bar J} Z + J {\bar Z} )} = 
{{l}\over{2\pi}} 
\prod_{n_1=-\infty}^{\infty} \prod_{n_2=0}^{\infty} 
\left ({{ C_{n_1n_2}}\over{2\pi}} \right ) \nonumber\\
&& \quad\quad \times \int d y_{n_1n_2} d{\bar y}_{n_1n_2}
 e^{- \quaft \left ( \sum_{n_1=-\infty}^{\infty}  \sum_{n_2=0}^{\infty}
   \omega_{n_1n_2} C_{n_1n_2} {\bar y}_{-n_1n_2} 
      y_{n_1n_2}\right ) } \nonumber\\
&& \quad\quad\quad \times e^{- 4 \pi \alpha^{\prime}
\sum_{n_1=-\infty}^{\infty} \sum_{n_2=0}^{\infty} 
\left [ (\omega_{n_1n_2} C_{n_1n_2})^{-1}
 \int_{M} d^2 \sigma d^2 \sigma^{\prime} {\bar J}(\sigma) J(\sigma^{\prime})
{\bar \Psi}_{-n_1n_2} (\sigma^{\prime}) \Psi_{n_1n_2} (\sigma) \right ]}
\quad ,
\label{eq:gf}
\end{eqnarray}
where the $y$ oscillators are defined by the shift:
\begin{eqnarray}
y_{n_1n_2} &&= z_{n_1n_2} - 
   {{i4\pi\alpha^{\prime}}\over{\omega_{n_1n_2}C_{n_1n_2}}}
\int d^2\sigma {\sqrt{g}} J {\bar \Psi}_{-n_1n_2}(\sigma) \nonumber \\
{\bar y}_{-n_1n_2} &&= {\bar z}_{-n_1n_2} - 
{{i4\pi\alpha^{\prime}}\over{\omega_{n_1n_2}C_{n_1n_2}}}
\int d^2\sigma {\sqrt{g}} {\bar J} \Psi_{n_1n_2}(\sigma) \quad .
\label{eq:shift}
\end{eqnarray} 
Substituting in the amplitude and performing the Gaussian integrations 
as before gives the result:
\begin{equation}
 {\cal A}(1, \cdots , N) = 
   \int_0^{\infty} {{dl}\over{l}} N_B [ \eta (\lh )]^{-24} 
e^{ - \sum_{i,j=1 }^{N} \sum_{m,n=0}^{(p-1)/2} 
   k^m_i {\bar k}^n_j G^{\prime}_{mn}(\sigma^a_i , \sigma^a_j ) 
 - \mu_1 \int d^2 \sigma {\sqrt{g}} } \quad .
\label{eq:mpoint}
\end{equation}
We include a term in the exponent proportional to the world-sheet 
cosmological constant in order to soak up any 
divergent contributions to the
Green's function. These will be understood as being absorbed in a
renormalization of the bare string coupling constant. 
We use a complex momentum basis with vectors,
$2 k_m $$=$$ p_{2m} $$+$$ i p_{2m+1}$, 
$2 {\bar k}_m $$=$$ p_{2m} $$-$$ i p_{2m+1}$, $m$$=$$0$, $\cdots$, $(p-1)/2$, 
and where the $p_{\mu}$ are real
momentum vectors. With this convention, the symmetric and 
antisymmetric bilinear products can be written as
\begin{eqnarray}
\half \left ( p_{2m}^i p_{2m}^j 
+ p_{2m+1}^i p_{2m+1}^j \right ) 
=&&  k^m_i {\bar k}^m_j + {\bar k}^m_i k^m_j   
\nonumber\\
{{i}\over{2}} \left ( p_{2m+1}^i p_{2m}^j 
- p_{2m+1}^j p_{2m}^i \right ) 
=&&  k^m_i {\bar k}^m_j - {\bar k}^m_i k^m_j   
\quad,
\label{eq:mom}
\end{eqnarray}
for each value of $m$. The 
Greens function for a complex scalar in the presence
of external sources takes the form
\begin{equation}
G^{\prime} (\sigma^a_i,\sigma^a_j)  = 
 \sum_{n_1=-\infty}^{\infty} \sum_{n_2=0}^{\infty} 
{{4 \pi \alpha^{\prime}}\over{\omega_{n_1n_2} C_{n_1n_2}}} 
    {\bar \Psi}_{-n_1n_2} (\sigma^a_i)\Psi_{n_1n_2} (\sigma^a_j) 
 \quad . 
\label{eq:green}
\end{equation}  
It satisfies Laplace's equation in the cylinder metric with 
the addition of a background charge given by the 
normalization of the zero mode derived above \cite{polbook}.
Note that the integration over zero modes, $z_0$, ${\bar z}_0$,
gives the usual delta function in momentum space in the large volume 
limit, and we have defined the normalization factor:
\begin{equation}
N_B = i (2\pi)^{p+1} g^N \delta ( \sum_{i=1}^{N} {\bf p}_i ) 
(4\pi^2 \alpha^{\prime} l )^{-(p+1)/2} {\rm det} ( {\bf 1} + {\bf B} ) 
\prod_{i=1}^N \int d s_i {\sqrt {g}} \quad .
\label{eq:norma}
\end{equation}
The derivation of the Greens function on the annulus is tedious, 
although straightforward, and is sketched in the appendix. For 
a single complex scalar and 
non-coincident sources, $i$$\neq$$j$, we obtain the result:
\begin{eqnarray}
G^{\prime} (\sigma^a_i,\sigma^a_j)  &&= - 
 {{2 \alpha^{\prime} }\over{1 + b^2}} 
  {\rm ln} {{|\Theta_{1}(\nu_{+}, \tau 
     ) |^2}\over{ |\eta( \tau ) |^2 }} 
- {{2i\alpha^{\prime}b }\over{1+b^2}}
 {\rm ln} {{\Theta_{1}(\nu_{+}, \tau ) 
    }\over{ \Theta_1^{\star}(\nu_{+},\tau)}}
 - \alpha^{\prime} {\rm ln} 
 | {{\Theta_{1}(\nu_{-}, \tau ) 
   }\over{ \Theta_{1}( \nu_{+},\tau )}} |^2 
\nonumber \\ && \quad 
+ {{2 \pi \alpha^{\prime} l}\over{1+b^2}}
 \left[\sigma^2 - {{2b\sigma}\over{l}}
 +{{2b^2}\over{3l^2}} \right ] 
\quad .
\label{eq:grbulkapp}
\end{eqnarray}
The notation is as follows:
\begin{equation}
 \nu_{\pm } = \half ( il \sigma_{-}^1 +  
    \sigma^2_{\pm} ) , \quad  
\tau = \lh  , \quad \sigma^a_{\pm} = 
\sigma^a_i \pm \sigma^a_j , \quad 
\sigma = \sigma_-^1 + {{b}\over{l}} \sigma^2_+  \quad . 
\label{eq:args}
\end{equation}
The Greens function for $(p+1)/2$ scalars coupled to distinct
constant background fields, $b_{(m)}$, $m$$=$$0$, $\cdots$,
$(p-1)/2$, is the direct sum of expressions of the form
given in Eq.\ (\ref{eq:grbulkapp}).\footnote{
An expression for the bulk Greens function 
has been derived using the background field method in \cite{gomis}.
Comparison with Eq.\ (3.8) of \cite{gomis} shows agreement for the first 
three terms in Eq.\ (\ref{eq:grbulkapp}). The derivation in \cite{gomis}
is missing the contributions from the constant modes, including a piece 
that cancels one of the terms present in Eq.\ (3.8) of Ref.\ \cite{gomis}. 
Notice that the term linear in $\sigma_{ij}$, which will contribute an 
external momentum dependent phase in the planar amplitude,
and which has opposite sign for the $\sigma^2$$=$$0$, $1$, boundaries, is 
absent in \cite{sang,bilal,gomis,multi}.}

The Greens function for two sources on the boundary of an annulus
can be straightforwardly
extracted from this expression. We will obtain the limiting
behavior of the Greens function for sources with small 
separation as follows. Setting 
$\sigma^2_i$$=$$\sigma^2_j$$=$$0$, $1$, gives the Greens functions
with two sources on the same boundary.
\begin{eqnarray}
G_P^{\prime} (\sigma_i,\sigma_j)  =&& - 
{{2 \alpha^{\prime} }\over{1 + b^2}}
  {\rm ln} {{|\Theta_{1}({{il\sigma_{ij}}\over{2}}, {{il}\over{2}} 
     ) |^2}\over{ |\eta( {{il}\over{2}} ) |^2 }}
+ {{2 \pi \alpha^{\prime} l}\over{1+b^2}}
   \left[(\sigma_{ij})^2 - {{2b\sigma_{ij}}\over{l}} +{{2b^2}\over{3l^2}}
\right ]
 + {{ 2 \pi \alpha^{\prime} b}\over{1+b^2}}
 {{\sigma_{ij}}\over{|\sigma_{ij}|}} 
\quad  \sigma^2=0 
\nonumber\\
G_P^{\prime} (\sigma_i,\sigma_j)  =&& - 
{{ 2 \alpha^{\prime} }\over{1 + b^2}}
  {\rm ln} {{|\Theta_{1}({{il\sigma_{ij}}\over{2}}, {{il}\over{2}} 
     ) |^2}\over{ |\eta( {{il}\over{2}} ) |^2 }}
+ {{2 \pi \alpha^{\prime} l}\over{1+b^2}}
   \left[(\sigma_{ij})^2 + {{2b\sigma_{ij}}\over{l}} +{{2b^2}\over{3l^2}}
\right ]
- {{ 2 \pi \alpha^{\prime} b}\over{1+b^2}}
 {{\sigma_{ij}}\over{|\sigma_{ij}|}} 
\quad  \sigma^2=1 
\quad.
\nonumber\\
\label{eq:grbndry}
\end{eqnarray}
We are henceforth suppressing the superscript
on $\sigma^1_-$$\equiv$$\sigma_{ij}$. 
We comment that the phases in these expressions have been 
determined directly
from the summations in the appendix, rather than from the 
compact expression for the bulk propagator given in Eq.\ (\ref{eq:grbulkapp}). 
Consequently, there is no ambiguity in the choice of branch for the 
logarithm function in the second term in Eq.\ (\ref{eq:grbulkapp}).

We now consider self-contractions. Since the Greens function is 
diagonal in the complex basis, it suffices to consider a single 
complex scalar.
The exponent in the expression for the N-point scattering amplitude 
given in Eq.\ (\ref{eq:mpoint})
can be rewritten as follows \cite{polbook}:
\begin{eqnarray}
\sum_{i,j=1 }^{N}   k_i {\bar k}_j 
   G^{\prime}_P (\sigma_i , \sigma_j )   
=&& \sum_{i \neq j;~ i,j=1 }^{N} k_i {\bar k}_j 
   G^{\prime}_P (\sigma_i , \sigma_j )   
   ~+~ \sum_{i=1 }^N  k_i {\bar  k}_i 
   G^{\prime}_P (\sigma_i , \sigma_i )   
\nonumber\\
=&&  
\half \sum_{i\neq j;~ i,j=1 }^{N} \left [ (  k_i {\bar  k}_j 
 +  k_j {\bar k}_i ) G^{\prime (e)}_P (\sigma_i , \sigma_j ) 
   + (  k_i {\bar k}_j - k_j {\bar k}_i ) G^{\prime (o)}_P 
      (\sigma_i , \sigma_j ) \right ] 
\nonumber\\
&& \quad -  
   \half \sum_{i \neq j; i,j=1 }^N ( k_i {\bar k}_j 
 +  k_j {\bar k}_i ) G^{\prime }_{P} (\sigma_i , \sigma_i )   
\quad ,
\label{eq:contract}
\end{eqnarray}
where we have used momentum conservation in obtaining the last step.
Note that we could further simplify this expression by writing the
double sums as summations over $i$, $j$, with restriction $i$$>$$j$,
or vice versa. The form given here will be found convenient in 
establishing, in section 4.1, the interpretation of the momentum 
dependent phase functions normalizing planar, and nonplanar, amplitudes 
as the wavefunction renormalization of open string vertex operators. 
Note that we have distinguished terms in the boundary Greens 
function that are even (odd) under the interchange of $i$ 
and $j$ by the superscripts $(e)$, $(o)$. The summation incorporates 
both the symmetric and the antisymmetric momentum bilinears defined 
in Eq.\ (\ref{eq:mom}). The presence in the Greens function of both 
the symmetric and antisymmetric invariants in the noncommutativity 
parameters, will carry through to the
momentum dependent phases normalizing the amplitude.

We focus on the self-contractions in the last term, obtained as a 
limit of the Greens function for closely separated sources. In the
zero $B$ field limit, this 
takes the general form:
\begin{equation}
\lim_{\sigma_i \to \sigma_j}
G^{\prime}_P (\sigma_i , \sigma_j ) = - 2 \alpha^{\prime} ~ {\rm ln} ~
d^2 (|\sigma_{ij}|)  
 + f(\sigma_i, \sigma_j) \quad ,
\label{eq:dist}
\end{equation}
where $d$ is the distance between the sources as measured on the 
world-sheet, and the function $f$ is finite in the limit 
$\sigma_i$$=$$\sigma_j$. 
Let us evaluate the leading short distance singularity in the Greens
function in the presence of a constant background $B$ field. 
In the limit of small separations, Eq.\ (\ref{eq:grbndry}) gives 
\begin{equation}
G_P^{\prime} (\sigma_i , \sigma_i ) = 
- {{ 2 \alpha^{\prime} }\over{1 + b^2}}
~ {\rm ln} ~ | l \sigma_{ij} |^2
- {{ 2 \alpha^{\prime} }\over{1 + b^2}}
~ {\rm ln} ~ | \pi (\eta(\lh))^2|^2
+ {{4 \pi \alpha^{\prime} b^2}\over{3l(1+b^2)}}
, \quad \sigma^2_i = 0, ~ 1 
 \quad .
\label{eq:distinv}
\end{equation}

The leading short distance divergence in Eq.\ (\ref{eq:distinv})
has the same form as on the 
boundary of the disk \cite{polbook,sw}, except that $d$ is specified 
with respect to the fiducial metric on the annulus. Note that one 
can define a regulated Greens function by subtracting the divergence:
\begin{equation}
G^{{\rm reg.}}_P(\sigma_i, \sigma_i) = G^{\prime}_P (\sigma_i, \sigma_i)
   + {{ 2 \alpha^{\prime} }\over{1 + b^2}}
        \lim_{\sigma_i \to \sigma_j}
 ~{\rm ln} ~ |l \sigma_{ij}|^2
\quad ,
\label{eq:regulate}
\end{equation}
on either boundary, making the same subtraction in Eqs.\ 
(\ref{eq:contract}) and (\ref{eq:green}). The regulated Greens 
functions enter into the derivation of the scattering amplitudes 
in the next section. The divergent terms are to be understood as 
having been absorbed in a renormalization of the bare open string 
coupling \cite{poltorus}.

Finally, we can write down the Greens function with sources on
different boundaries:
\begin{equation}
G_{NP}^{\prime} (\sigma_i,\sigma_j)  = - 
{{2 \alpha^{\prime} }\over{1 + b^2}}
  {\rm ln} | {{\Theta_{2}({{il\sigma_{ij}}\over{2}}, {{il}\over{2}} 
     )}\over{ \eta( {{il}\over{2}} ) }} |^2
+ {{2 \pi \alpha^{\prime} l}\over{1+b^2}}
   \left[(\sigma_{ij})^2 - {{b^2}\over{3l^2}} \right]  
\quad\quad \sigma^2_+=1 \quad .
\label{eq:gnonp}
\end{equation}
Notice that the Greens function with sources on different boundaries 
is even under their interchange and, as expected, there are no short 
distance divergences.

\section{Wavefunction Renormalization and Planar Amplitudes}
\label{sec:scatter}

Consider now the planar amplitude with $N$ tachyon vertex operator
insertions on a single boundary of the annulus. The amplitude can 
be put in a more transparent form as follows. Following 
\cite{sw}\cite{schomerus}\cite{srm}, it is natural to identify symmetric,
and antisymmetric, momentum bilinears for any given pair of vertex operators,
$(V_i,V_j)$, with {\em complex} momenta, $({\bf k}_i,{\bf k}_j)$:
\begin{eqnarray}
  {\bf k}_i \cdot {\bar {\bf k}}_j =&& 
\sum_{m=0}^{(p-1)/2} 
{{ \alpha^{\prime} }\over{1 + b_{(m)}^2}}
( k^{m}_i {\bar k}^m_{j} + k^{m}_j {\bar k}^m_{i} )  
\nonumber \\
  {\bf k}_i \circ {\bar {\bf k}}_j =&& 
\sum_{m=0}^{(p-1)/2} 
{{ \pi \alpha^{\prime} b^2_{(m)} }\over{1 + b_{(m)}^2}}
( k^{m}_i {\bar k}^m_{j} + k^{m}_j {\bar k}^m_{i} )  
\nonumber \\
    {\bf k}_i \times {\bar {\bf k}}_j =&&  \sum_{m=0}^{(p-1)/2} 
{{2 \pi \alpha^{\prime} b_{(m)} }\over{1+b_{(m)}^2}} 
      ( k^{m}_i {\bar k}^m_{j} 
      - k^{m}_j {\bar k}^m_{i} ) 
= \sum_{m=0}^{(p-1)/2} 
 \theta_{(m)} 
      ( k^{m}_i {\bar k}^m_{j} 
      - k^{m}_j {\bar k}^m_{i} ) 
 \quad .
\label{eq:prods}
\end{eqnarray}
Note that one can construct two distinct symmetric bilinear
products in the momentum by taking appropriate powers of $B$. Only the first 
appears in the planar amplitude.
The antisymmetric bilinear product accompanies terms in the Greens 
function that are odd under the interchange of $i$ with $j$.
Both symmetric and antisymmetric forms appear in the 
scattering amplitudes, and are also
present in the vertex operator algebra of string theory in a background
$B$ field. The spacetime invariants implicit in the bilinear products
given in Eq.\ (\ref{eq:prods}) are manifest in the alternative basis:
\footnote{Note that ${\bf k}
\circ {\bar{\bf k}}$ is identical in form to the spacetime invariant, 
${\bf p} \circ {\bf p}$, appearing in the noncommutative field 
theory analysis of \cite{srm}. However, it is {\em dimensionless}.} 
\begin{eqnarray}
{\bf k}_i \cdot {\bar {\bf k}}_j =&& 
{{\alpha^{\prime}}\over{2}}
\sum_{\mu,\nu=0}^{p}
 ( {\cal M}^{-1} )^{\mu\nu} p^i_{\mu} p^j_{\nu} 
\nonumber\\
{\bf k}_i \circ {\bar {\bf k}}_j =&& 
 - { {1}\over{8\pi \alpha^{\prime} } } 
\sum_{\mu,\nu=0}^{p}
 \Theta^{\mu\sigma} ( {\cal M})_{\sigma\lambda} \Theta^{\lambda\nu} 
   ~ p^i_{\mu} p^j_{\nu} 
\nonumber \\
{\bf k}_i \times {\bar {\bf k}}_j =&& -{{i}\over{2}}  
\sum_{\mu,\nu=0}^{p}
\Theta^{\mu\nu} ~ p^i_{\mu} p^j_{\nu} 
 \quad ,
\label{eq:pprods}
\end{eqnarray}
where spacetime indices are raised and lowered using the metric,
$g_{\mu\nu}$$=$$\delta_{\mu\nu}$, and ${\cal M}^{-1}$,
$\Theta_{\mu\nu}$ are defined by the spacetime commutation 
relations given in Eq.\ (\ref{eq:comm}). Note that the 
first of the spacetime invariants can be identified 
with the \lq\lq open string 
metric" of \cite{sw}. We have used Eq.\ (\ref{eq:mom})
to relate the real and complex momentum bases. The field theory 
limit corresponds to taking $\alpha^{\prime}$ to
zero, keeping fixed the dimensionless variables, $B$, ${\cal M}$,
and $\Theta/\alpha^{\prime}$.\footnote{This does {\em not}
coincide with the scaling limit described in \cite{sw} since,
with our conventions, $B$ is dimensionless.}

It is convenient to perform all calculations in the complex
basis. Substituting for the Greens function 
in Eq.\ (\ref{eq:green}), the one-loop planar amplitude 
takes the form:
\begin{eqnarray}
{\cal A}_P (1, \cdots , N) =&&  
i {\tilde g}^N \delta ( \sum_{i=1}^{N} {\bf p}_i ) 
(\alpha^{\prime})^{-(p+1)/2} {\rm det} ( {\bf 1} + {\bf B} ) 
\left [ \prod_{r=1}^{N} \int d \sigma_r \right ] 
\int_0^{\infty} {{dl}\over{l}}
e^{-\mu_2 \int d^2 \sigma {\sqrt{g}} }
\nonumber \\
&& \quad \times 
{{l^{N-(p+1)/2}}\over{ \eta(\lh)^{24}}}
 \prod_{i \neq j;~ i,j=1}^{N} \left [
e^{ {\bf k}_i \times {\bar {\bf k}}_j \left ( \sigma_{ij}  
- \half {{\sigma_{ij}}\over{|\sigma_{ij}|}} \right ) } 
f_1(l; \sigma_{ij}) \right ]
\quad ,
\label{eq:formpr}
\end{eqnarray}
where we have defined the functions:
\begin{equation}
f_a(l;\sigma) \equiv 
e^{ - \pi l \sigma_{ij}^2  {\bf k}_i \cdot {\bar {\bf k}}_{j} 
}
| {{ \Theta_{a}({{il}\over{2}}\sigma_{ij},{{il}\over{2}} )}\over
    {\pi\eta^3 ({{il}\over{2}})}}|^{ 2 {\bf k}_i \cdot {\bar {\bf k}}_{j}}
\quad ,
\label{eq:fli}
\end{equation}
and included a term in the exponent proportional to the world-sheet
cosmological constant in order to soak up any divergent terms generated in
the renormalized amplitude. 
We have used the form of the boundary propagator 
given in 
the first of Eqs.\ (\ref{eq:grbndry}). Notice that the difference
in the sign of the odd part of the Greens function is correlated 
with the ordering of the sources on the boundary.
Note also that the bare open string coupling, $g$, has been 
replaced by the multiplicatively renormalized string 
coupling, ${\tilde{g}}$, \cite{poltorus} as discussed in
the previous section.
The pre-factor linear in $\sigma_{ij}$ can be recognized 
using Eq.\ (\ref{eq:pprods}):
\begin{equation}
{\rm exp} \left [ ( {\bf k}_i \times {\bar {\bf k}}_j ) \sigma_{ij}  \right ]
= {\rm exp} \left [ 
- {{i}\over{2}} \Theta^{\mu\nu} p^i_{\mu} p^j_{\nu}  \sigma_{ij} \right ]
\quad .
\label{eq:star}
\end{equation}
The star product, denoted by the $\times$ symbol in Eqs.\ (\ref{eq:pprods}), 
makes a natural 
appearance in the antisymmetric bilinear product of the external momenta.

Consider now the field theory limit in which only the massless open 
string mode propagates around the loop. This is extracted from
the expression above by isolating the $O(q^0)$ term in an
expansion of the integrand in powers of $q$$=$$e^{-\pi l}$:
\begin{eqnarray}
&&\lim_{l\to \infty} \left \{ \eta^{-24} 
~ | {{ \Theta_{1}({{il}\over{2}}\sigma_{ij},{{il}\over{2}} )}\over
    {\pi\eta^3 ({{il}\over{2}})}}|^{ 2 {\bf k}_i \cdot {\bar {\bf k}}_{j} } \right \} 
\nonumber\\
&& \quad = e^{\pi l} \left ( 1+ 24 e^{-\pi l} \right ) 
~ |{{2}\over{\pi}} {\rm Sinh}(\half\pi l\sigma_{ij})|^{ 2 
   {\bf k}_i \cdot {\bar {\bf k}}_{j} } 
\left ( 1 - [8 {\bf k}_i \cdot {\bar {\bf k}}_{j}{\rm Sinh}^2(\half \pi l \sigma_{ij})] e^{-\pi l} \right ) 
\quad .
\label{eq:lims}
\end{eqnarray}
upto $O(e^{-\pi l})$ corrections. We suppress the leading term of 
$O(e^{\pi l})$, which corresponds
to the open string tachyon propagating in the loop, focusing 
on the $O(1)$ correction. 
In this limit, the scattering amplitude takes the form:
\begin{eqnarray}
{\cal A}_P |_{{\rm massless}} &&= 
  i {\tilde g}^N \delta ( \sum_{i=1}^{N} {\bf p}_i ) 
(\alpha^{\prime})^{-(p+1)/2} {\rm det} ( {\bf 1} + {\bf B} ) 
\left [ \prod_{r=1}^{N} \int d \sigma_r \right ] 
  \int_0^{\infty} dl l^{N-(p+3)/2}
e^{-\mu_2 \int d^2 \sigma {\sqrt{g}} }
 \nonumber \\
&& \quad \quad \times 
\prod_{i \neq j;~ i,j=1}^{N}
e^{ ({\bf k}_i \times {\bar{\bf k }}_j) 
  ( \sigma_{ij} - \half {{\sigma_{ij}}\over{|\sigma_{ij}|}} )  
 }
     \{ e^{ - \pi l \sigma_{ij}^2 {\bf k}_i \cdot {\bar {\bf k}}_{j} }
~ | {{2}\over{\pi}} {\rm Sinh}( \half \pi l 
   \sigma_{ij})|^{ 2 {\bf k}_i \cdot {\bar {\bf k}}_{j} } 
\nonumber \\
&& \quad\quad\quad\quad  \times 
~ \left 
[ 24 - 8 {\bf k}_i \cdot {\bar {\bf k}}_{j} 
{\rm Sinh}^2(\half\pi l \sigma_{ij})  \right ] \}
\quad .
\label{eq:formprlim}
\end{eqnarray}
It should be emphasized that the corrections from massive open string states
circulating in the loop do not change the 
basic form of this result, the single term given here being replaced by a series of the form:
\begin{equation} 
\sum_{n=0}^{\infty} ~ F^{({\rm open})}_n \left ( {\rm Sinh}(\half\pi l \sigma_{ij}) \right ) 
    e^{-(\sigma^2_{ij}+n)\pi l}
\quad ,
\label{eq:poly}
\end{equation}
where $F^{({\rm open})}_n$ is a polynomial function of the 
${\rm Sinh}(\half\pi l \sigma_{ij}) $, $i$,$j$$=$$1$, $\cdots$, $N$. The 
$n$th term in the series represents the contribution from states in the 
$n$th mass level of the open string spectrum. 
The effective string tension introduced earlier defines a natural 
scale with respect to which we can measure external momentum. We will 
now restrict to on-shell external momenta setting 
\begin{equation}
{\bf k}_i \cdot {\bar {\bf k}}_i =
\sum_{m=0}^{(p-1)/2} 
 {{2 \alpha^{\prime} }\over{1+b_{(m)}^2}}
    k^m_i {\bar k}^m_{i}  = {{n_i}\over{2}}
\quad ,
\label{eq:onshell}
\end{equation}
where the $n_i$ take integer values. Vertex operators with $n_i$$=$$1$ 
correspond to on-shell open string tachyons, with the delta function imposing 
momentum conservation. In contrast to the field theory limit of the planar 
amplitudes of free string theory in zero background $B$ field and ordinary 
commutative space, the noncommutative amplitude is characterized
by the presence of exponentiated phases:
\begin{equation}
e^{ ({\bf k}_i \times {\bar{\bf k }}_j) \left ( \sigma_{ij} 
   - \half {{\sigma_{ij}}\over{|\sigma_{ij}|}} \right )  
 } 
= e^{- \half \sum_{\mu,\nu=0}^{p} 
i \Theta^{\mu\nu} p^i_{\mu} p^j_{\nu}
  \left ( \sigma_{ij} 
   - \half {{\sigma_{ij}}\over{|\sigma_{ij}|}} \right )  
  } 
\quad ,
\label{eq:phases}
\end{equation}
for all $i$, $j$$=$$1$, $\cdots$, $N$. The antisymmetric star products of 
the external momenta make a natural appearance, in accordance with the 
expectations of noncommutative scalar field theory \cite{flk}.
In order to make a more direct comparison with the results of 
noncommutative field theory, it is helpful to explore further
the external momentum dependent normalization of the planar amplitude.

A natural interpretation of a momentum dependent phase function normalizing
the planar amplitude is that it represents a wavefunction renormalization. 
To ascertain the form of such a phase function it is helpful to carry out the 
integrations over the modular parameter, $l$, and the $\sigma_i$ explicitly, 
making a suitably simple choice of kinematics.\footnote{An alternative 
treatment of the $l$$\to$$\infty$ limit focuses on 
the $O(1)$ term in the $e^{-\pi/l}$ expansion without performing the 
integration
over the modular parameter. Upon applying momentum conservation, the 
integrals over the $\sigma_i$ are performed, 
leaving the result in the form of
an integral representation valid for generic kinematics \cite{grsch}. The 
method could also be applied here.} We will show now that upon imposing the 
on-shell conditions on the external 
momenta, and in the restricted kinematics of forward scattering,
the integrals over the modular parameter, $l$, and the $\sigma_i$ can 
in fact be performed explicitly. The result 
clarifies that the momentum dependent 
phase function normalizing the amplitude obtained as a result of both 
integrations, can indeed be interpreted as a wavefunction 
renormalization of the open string vertex operators.

\subsection{Forward Scattering and the Decoupling Limit}
\label{sec:kinem}

Consider the limit of forward 
scattering with ${\rm cos}(\phi_{ij})$$=$$\pm 1$, 
for every $i$,$j$$=$$1$, $\cdots$, $N$, in addition to the on-shell 
conditions imposed in Eq.\ (\ref{eq:onshell}). To accommodate momentum 
conservation with zero momentum transfer, $\sum_{i=1}^N {\bf p}_i$$=$$0$,
and finite, even, $N$, we could, for example, make the 
symmetric choice of $N/2$ parallel incoming and $N/2$ parallel outgoing on-shell 
tachyons. Alternatively, we could consider $N$$-$$1$ parallel incoming on-shell 
tachyons, 
and a single outgoing off-shell tachyon with 
$2{\bf k}_N \cdot{\bar {\bf k}}_N 
$$=$$N$$-$$1$. For either choice, a glance at Eq.\ (\ref{eq:formprlim})
shows that the integration over $l$ can be performed explicitly. The $\sigma_i$
integrals can then be performed by the iterative method described below.
However, it must be kept in mind that zero momentum transfer implies the
regime, ${\bf p}^2$$<<$$\alpha^{\prime}_{{\rm eff.}}$, which is properly
examined in the closed string channel, as in the next section. We will
instead examine here the
forward scattering of $N$ on-shell tachyons, where $N$ is a large number. 
Momentum conservation implies that the forward momentum is 
carried away by the closed string state which is, consequently, infinitely 
massive for sufficiently large $N$. In the limit of large momentum transfer, 
the decoupling of the massive closed string states from the massless states 
in the field theory limit of the open string sector is natural, and the 
restriction to the $O(1)$ term in the expansion in $e^{-\pi l}$ increasingly 
well-justified. In the limit of infinite momentum transfer, the approximation 
becomes exact. We will therefore consider the forward scattering 
of $N$ parallel 
on-shell tachyons as a simple and concrete realization of the planar amplitude 
in the decoupling limit. In section 5, we will consider the opposite
regime of zero momentum transfer.

Setting 
$2 {\bf k}_i \cdot{\bar {\bf k}}_j$$=$$1$ for all $i$, $j$$=$$1$, $\cdots$,
$N$, the planar amplitude takes the simple form:
\begin{eqnarray}
{\cal A}_P |_{{\rm decoup.}} &&= 
i {\tilde g}^N \delta ( \sum_{i=1}^{N} {\bf p}_i - {\bf p}) 
  \prod_{i,j=1}^N \delta ({\bf p}_i \cdot {\bf p}_j -1)
(\alpha^{\prime})^{-(p+1)/2} {\rm det} ( {\bf 1} + {\bf B} ) 
\left [ \prod_{r=1}^{N} \int d \sigma_r \right ] 
 \nonumber \\
&& \quad\quad  \times 
  \int_0^{\infty} dl l^{N-(p+3)/2}
e^{-\mu_2 \int d^2 \sigma {\sqrt{g}} }
\prod_{i \neq j;~ i,j=1}^{N}
e^{ ( {\bf k}_i \times {\bar{\bf k}}_j )\left (\sigma_{ij}  
   - \half {{\sigma_{ij}}\over{|\sigma_{ij}|}} \right )  
 } 
 \nonumber \\
&& \quad\quad\quad \quad  \times 
     \left \{ e^{ - \half \pi l \sigma_{ij}^2  }
2^{4} \pi^{-1} 
\left ( 3 ~ {\rm Sinh} ( \half \pi l \sigma_{ij}) - \half
    {\rm Sinh}^2 (\half\pi l \sigma_{ij})  \right ) 
\right \}
\quad ,
\nonumber\\
\label{eq:formlim}
\end{eqnarray}
where ${\bf p}$ is the net forward momentum transfer, understood to be
absorbed by a massive closed string state. Notice that the world-sheet 
cosmological constant term regulates the large $l$ behavior of the 
integral, any divergences being absorbed in a
renormalization of the string coupling constant. The result can be 
expressed in the form:
\begin{eqnarray}
{\cal A}_P |_{{\rm decoup.}} =&&
i {\tilde g}^N \delta ( \sum_{i=1}^{N} {\bf p}_i - {\bf p}) 
  \prod_{i,j=1}^N \delta ({\bf p}_i \cdot {\bf p}_j -1)
(\alpha^{\prime})^{-(p+1)/2} {\rm det} ( {\bf 1} + {\bf B} ) 
\left [ \prod_{r=1}^{N} \int d \sigma_r \right ] 
\nonumber\\
\quad\quad\quad\quad &&\times\prod_{i \neq j;~ i,j=1}^{N}
e^{ ( {\bf k}_i \times {\bar{\bf k}}_j )\left (\sigma_{ij}  
   - \half {{\sigma_{ij}}\over{|\sigma_{ij}|}} \right )  
} 
    P (\sigma_1, \sigma_2, \cdots , \sigma_N)
\quad ,
\label{eq:resp}
\end{eqnarray}
where $P$ is a polynomial function of the $\sigma_i$, $i$$=$$1$, 
$\cdots$, $N$, of order $n_p$$=$${{(p+1)}\over{2}}$$-$$N$, 
which can be expressed in the form:
\begin{equation}
P (\sigma_1, \sigma_2, \cdots , \sigma_N)
= \sum_{m_N=0}^{n_p} \cdots \sum_{m_2=0}^{n_p} 
  \sum_{m_1=0}^{n_p} \left ( a^{(1)}_{m_1m_2 \cdots m_N} 
  \sigma_{N}^{m_N} \cdots \sigma_2^{m_2} \right ) \sigma_{1}^{m_1}
\quad .
\label{eq:polys}
\end{equation}
Note that with the chosen ordering of sources on the boundary,
we have $\sigma_{ij}$$>$$0$ for every $i$$>$$j$, with the 
integration domains, $0$$\le$$\sigma_1$$\le$$\sigma_{2}$, 
$0$$\le$$\sigma_2$$\le$$\sigma_{3}$, $\cdots $, and 
$0$$\le$$\sigma_N$$\le$$1$. The nested set of integrals over
the $\sigma_i$ can be evaluated as follows.

It is convenient to define the momentum dependent phases
\begin{equation}
\phi_j \equiv  - \sum_{i \neq j, i=1}^N 
   ( {\bf k}_i \times {\bar {\bf k}}_j ) 
= {{i}\over{2}} \sum_{i \neq j, i=1}^N \Theta^{\mu\nu} p^i_{\mu} p^j_{\nu} 
\quad ,
\label{eq:phase}
\end{equation}
in terms of which the innermost integral in Eq.\ (\ref{eq:resp})
takes the form:
\begin{eqnarray}
I_1 =&& \int_0^{\sigma_{2}} d\sigma_1
 e^{ -\phi_1  \sigma_{1} + R_1(\sigma_2, \cdots ,\sigma_{N}) } 
  \sum_{m_{1}=0}^{n_p} \left( a^{(1)}_{m_1 m_2 \cdots m_N} 
         \sigma_{N}^{m_N} \cdots \sigma_2^{m_2} \right ) \sigma_{1}^{m_1}
\nonumber\\
    =&& e^{-\phi_2\sigma_2 + R_2(\sigma_3, \cdots ,\sigma_{N}) } 
                  \sum_{m_{1}=0}^{n_p} \left( a^{(1)}_{m_1m_2 \cdots m_N} 
                     \sigma_{N}^{m_N} \cdots \sigma_3^{m_3} \right ) \sigma_{2}^{m_2}
                       {\bf \gamma} \left ( m_1+1,\phi_1\sigma_2 \right ) 
\quad,
\label{eq:nestone}
\end{eqnarray}
where ${\bf \gamma}$ is an incomplete gamma function, and the
$R_j$ are linear in the $\sigma_i$, $i$$>$$j$. Substituting its
polynomial representation, we iterate the procedure above to perform
the next-to-innermost integral over $\sigma_2$:
\begin{eqnarray}
I_2 =&& \int_0^{\sigma_{3}} d\sigma_2
                e^{ R_2(\sigma_3, \cdots ,\sigma_{N}) } 
                    \sum_{m_2=0}^{n_p} \sum_{m_1=0}^{n_p} 
                         \left ( a^{(1)}_{m_1m_2 \cdots m_N} 
                 \sigma_N^{m_N} \cdots \sigma_3^{m_3} \right ) 
                    \left [ (m_1)! \right ] 
\nonumber\\
 && \quad\quad
          \left \{ 
               e^{-\phi_2 \sigma_2  } \sigma_{2}^{m_2}
                   -  e^{-(\phi_1 + \phi_2 ) \sigma_2  } 
                     \left ( \sum_{r_1=0}^{m_1} {{\phi_1^{r_1}}\over{r_1!}} \sigma_{2}^{r_1+ m_2}
                        \right ) 
                 \right \}
\nonumber\\
            =&&  e^{ R_2 (\sigma_3 , \cdots , \sigma_N ) } 
                    \sum_{m_2 =0}^{n_p} \sum_{m_1 =0}^{n_p} 
                         \left ( a^{(1)}_{m_1m_2 \cdots m_N} 
                 \sigma_{N}^{m_N} \cdots \sigma_3^{m_3} \right ) 
                    \left [ (m_1)! \right ] 
\nonumber\\
&&\quad\quad
                  \times   \left \{ {\bf \gamma} \left ( m_2+1,\phi_2\sigma_3 \right ) 
               - \left [ \sum_{r_1=0}^{m_1} {{\phi_1^{r_1}}\over{r_1!}} 
                    {\bf \gamma} \left ( r_1+m_2+1,(\phi_1+\phi_2)\sigma_3 \right ) \right ] 
                   \right \}
\nonumber\\
            =&&  e^{ R_3 (\sigma_4 , \cdots , \sigma_N ) } 
                    \sum_{m_2 =0}^{n_p} \sum_{m_1 =0}^{n_p} 
                         \left ( a^{(1)}_{m_1m_2 \cdots m_N} 
                 \sigma_{N}^{m_N} \cdots \sigma_3^{m_3} \right ) 
                    \left [ \prod_{i=1,2} (m_i)! \right ] 
\nonumber\\
&&\quad\quad
                     \times \{  e^{-\phi_3 \sigma_3  } \sigma_{3}^{m_3}
                              -  e^{-(\phi_2 + \phi_3 ) \sigma_3  } 
                     \left ( \sum_{r_2=0}^{m_2} {{\phi_2^{r_2}}\over{r_2!}} \sigma_{3}^{r_2+ m_3}
                        \right ) 
               -  e^{-\phi_3\sigma_3 } \sum_{r_1=0}^{m_1} 
                    {{(r_1+m_2)!}\over{m_2!}} {{\phi_1^{r_1}}\over{r_1!}} \sigma_3^{m_3} 
\nonumber\\
&&\quad\quad\quad
                             - e^{-(\phi_1+\phi_2+\phi_3)\sigma_3 } 
                                  \sum_{r_2=0}^{m_2} \sum_{r_1=0}^{m_1} 
                               {{(r_1+m_2)!}\over{m_2!}}
                                   {{(\phi_1+\phi_2)^{r_1+m_2+r_2}}\over{(r_1+m_2+r_2)!}}
                                        \sigma_3^{r_1+m_2+r_2+m_3}
                       \}
\quad,
\label{eq:nesttwo}
\end{eqnarray}
continuing through to the $N$th integral. The result takes the form:
\begin{eqnarray}
I_N =&& 
\sum_{m_N=0}^{n_p} \cdots \sum_{m_1=0}^{n_p} a^{(1)}_{m_1m_2 \cdots m_N} 
    \{ q(\phi_i; i\le N)  - 
\sum_{j=1}^N \sum_{r_j=0}^{m_j} e^{-\phi_j  } q_j(\phi_i; i \le N) 
+ \cdots 
\nonumber\\
&& \quad \quad + e^{- (\phi_1 + \phi_2 + \cdots + \phi_N)}
\sum_{r_N=0}^{m_N} \cdots \sum_{r_1=0}^{m_1} 
                    q_{1 \cdots N} (\phi_i; i \le N)
 \}
\nonumber\\
\equiv&& e^{- (\phi_1 + \phi_2 + \cdots + \phi_N)} 
Q(\phi_1, \cdots , \phi_N)
\quad .
\label{eq:nest}
\end{eqnarray}
where the $q$'s are polynomial functions of the $\phi_i$. For particular values of 
$p$, $N$, this procedure can be made explicit inclusive of numerical factors.
Thus, in contrast to the field theory limit of the planar amplitudes of free string 
theory in zero background $B$ field and ordinary commutative space, the noncommutative 
amplitude is normalized by a momentum dependent phase function. We have shown 
that the functional dependence of the external momentum dependent
phase function normalizing the planar amplitude takes the form:
\begin{eqnarray}
{\cal A}_P |_{{\rm decoup.}} =&&
i {\tilde g}^N \delta ( \sum_{i=1}^{N} {\bf p}_i - {\bf p}) 
  \prod_{i,j=1}^N \delta ({\bf p}_i \cdot {\bf p}_j -1)
(\alpha^{\prime})^{-(p+1)/2} {\rm det} ( {\bf 1} + {\bf B} ) 
\nonumber\\
\quad\quad\quad && \times
e^{- {{1}\over{4}} \sum_{\mu,\nu=0}^p \sum_{j=1}^N \sum_{i \neq j; i=1}^N
 i \Theta^{\mu\nu} p^i_{\mu} p^j_{\nu}  
} 
\nonumber\\
\quad\quad\quad\quad\quad &&\times 
Q \left ( \phi_1,\phi_2, \cdots , \phi_N \right )
\quad ,
\label{eq:copole}
\end{eqnarray}
where we have substituted for the phases, $\phi_i$, 
$i$$=$$1$, $\cdots$, $N$, defined in 
Eq.\ (\ref{eq:phase}). Note that this expression holds 
for a chosen ordering  of the external momenta about the
boundary. The delta functions impose 
momentum conservation and the on-shell condition for 
the external momenta.

It is now apparent that the momentum dependent phase 
function has a natural
interpretation as a wavefunction renormalization. Note that the result 
is a function of the variables, $ \sum_{i\neq j; i=1}^N ({\bf p}_i \times 
{\bf p}_j) $, for every $j$$=$$1$, $\cdots$, $N$. Thus, the $j$th 
vertex operator with momentum ${\bf p}_j$ is normalized by a $B$
dependent phase function determined by ${\bf p}_j$, and
antisymmetric bilinear products coupling each ${\bf p}_j$ to the 
total remnant loop momentum due to the other sources: 
${\bf p}$$-$${\bf p}_j$$=$$\sum_{i\neq j} {\bf p}_i$.

\section{World-sheet Duality and Non-planar Amplitudes}
\label{sec:np}

The non-planar amplitude, with $N_1$ vertex operators on one 
boundary of the annulus and $N$$-$$N_1$ vertex operators on 
the other, takes the form:
\begin{eqnarray}
 {\cal A}_{NP} (1, \cdots , M)&&=
i {\tilde g}^N 
\delta \left ( \sum_{a=1}^{N} {\bf p}_a \right ) 
(\alpha^{\prime})^{-(p+1)/2} {\rm det} ( {\bf 1} + {\bf B} ) 
\left [ \prod_{a=1}^{N} \int d \sigma_a \right ]
\nonumber \\
&& \quad\quad \times 
   \int_0^{\infty} {{dl}\over{l}} 
{{ l^{N-(p+1)/2}}\over{ \eta ( {{il}\over{2}} )^{24} }}
\left [
e^{ - \sum_{i,j=1 }^{N_1} \sum_{m=0}^{(p-1)/2} 
   k^m_i {\bar k}^m_j G^{\prime}_P(\sigma^a_i , \sigma^a_j ) }  
\right ]
\nonumber \\
&& \quad\quad\quad \times 
\left [ 
e^{ - \sum_{k=1}^{N_1} 
\sum_{u=N_1+1}^{N} 
\sum_{m=0}^{(p-1)/2} 
   ( k^m_k {\bar k}^m_u +  k^m_u {\bar k}^m_k ) 
G^{\prime}_{NP}(\sigma^a_k , \sigma^a_u ) }  
\right ]
\nonumber \\
&& \quad\quad\quad\quad \times 
\left [ 
e^{ - \sum_{r,t=N_1+1 }^{N} \sum_{m=0}^{(p-1)/2} 
   k^m_r {\bar k}^m_t G^{\prime}_P (\sigma^a_r , \sigma^a_t ) }  
\right ]
\quad .
\nonumber\\
\label{eq:nonpf}
\end{eqnarray}
The derivation mirrors the earlier treatment of the planar amplitude.
The Greens function for a single complex scalar with $N_1$ sources on the 
$\sigma^2$$=$$0$ boundary and $\sum_{i=1}^{N_1} {\bf p}_i $$=$${\bf p}$,
can be expressed in the form:
\begin{eqnarray}
\sum_{i,j=1 }^{N_1}   k_i {\bar  k}_j 
   G^{\prime}_P (\sigma_i , \sigma_j )   
=&& \sum_{i\neq j;~ i,j=1 }^{N_1} k_i {\bar k}_j 
   G^{\prime }_P (\sigma_i , \sigma_j )   
   + \sum_{i=1 }^{N_1}  k_i {\bar  k}_i 
   G^{\prime}_P (\sigma_i , \sigma_i )   
\nonumber\\
=&& \sum_{i\neq j;~ i,j=1 }^{N_1}  k_i {\bar k}_j 
   G^{\prime }_P (\sigma_i , \sigma_j )   
   ~+~ \sum_{i=1 }^{N_1}  k_i {\bar k} 
   G^{\prime}_P (\sigma_i , \sigma_i )   
   ~ - \sum_{i\neq j;~ i,j=1 }^{N_1} k_i {\bar  k}_j 
   G^{\prime}_P (\sigma_i , \sigma_i )   
\nonumber\\
=&& \half \sum_{i \neq j;~ i,j=1 }^{N_1} \left [ ( k_i {\bar k}_j 
 +  k_j {\bar  k}_i ) 
   G^{\prime (e)}_P (\sigma_i , \sigma_j )   
+ (  k_i {\bar  k}_j - k_j {\bar k}_i ) 
   G^{\prime (o)}_P (\sigma_i , \sigma_j )   
\right ]
\nonumber\\
   && \quad\quad
   + ~ \sum_{i=1 }^{N_1}  k_i {\bar k} 
   G^{\prime}_P (\sigma_i , \sigma_i )   
 ~-~ \half \sum_{i \neq j; i,j=1 }^N 
(  k_i {\bar  k}_j +  k_j {\bar k}_i ) 
  G^{\prime}_P (\sigma_i , \sigma_i )   
\quad ,
\nonumber\\
\label{eq:contractnp}
\end{eqnarray}
and likewise for sources on the $\sigma^2$$=$$1$ boundary,
but with $k$ replaced by $-k$. The terms dependent on
the internal momentum can be rewritten as the unrestricted 
double sums:
\begin{eqnarray}
\sum_{i=1 }^N \sum_{r=N_1+1 }^N 
  k_i (-{\bar k}_r) 
  G^{\prime}_P (\sigma_i , \sigma_i )   
 \quad \quad &&{\rm for} ~ \sigma^2_i = 0 
\nonumber\\
- \sum_{r=N_1+1 }^N \sum_{j=N_1 }^N 
  k_r (+{\bar k}_j) 
  G^{\prime}_P (\sigma_r , \sigma_r )   
 \quad\quad  &&{\rm for} ~ \sigma^2_r = 1 \quad ,
\label{eq:sums}
\end{eqnarray}
combining neatly with the unrestricted double sums over
propagators, $G_{NP}^{\prime}(\sigma_i,\sigma_r)$, between 
sources on distinct boundaries. Substituting from Eqs.\ 
(\ref{eq:distinv}) and (\ref{eq:gnonp}) in Eq.\ 
(\ref{eq:green}) gives the expression:
\begin{equation}
 \prod_{i=1}^{N_1} \prod_{r=N_1+1}^N 
e^{{{2 \pi \alpha^{\prime} b^2}\over{l(1+b^2)}} 
({\bf k}_i {\bar {\bf k}}_r + {\bf k}_r {\bar {\bf k}}_i ) } 
 \left [ e^{ - 2\pi l \sigma_{ir}^2 {\bf k}_i \cdot {\bar {\bf k}}_{r} 
}
| {{ \Theta_{2}({{il}\over{2}}\sigma_{ir},{{il}\over{2}} )}\over
    {\pi \eta^3 ({{il}\over{2}})}}|^{ 4 {\bf k}_i \cdot {\bar {\bf k}}_{r}}
  \right ] \equiv \prod_{i=1}^{N_1} \prod_{r=N_1+1}^N f^2_2(l;\sigma_{ir}) 
                      e^{ 2 {{{\bf k}_i \circ {\bar {\bf k}}_r}\over{l}} }
\quad ,
\label{eq:fl}
\end{equation}
Here, ${\bf k}$$=$$\sum_{i=1}^{N_1} 
{\bf k}_i $$=$$-\sum_{r=N_1+1}^N
{\bf k}_r$ is the momentum transfer between the boundaries.
Notice the appearance in the nonplanar amplitude of the 
symmetric product defined in Eq.\ (\ref{eq:prods}), corresponding to
the spacetime invariant \cite{sw,srm}:
\begin{equation}
\sum_{i=1}^{N_1}\sum_{r=N_1+1}^N  {\bf k}_i \circ {\bar {\bf k}}_r
= - {\bf k} \circ {\bar {\bf k}}
= { {1}\over{8\pi \alpha^{\prime} } } 
\sum_{\mu,\nu=0}^{p}
 \Theta^{\mu\sigma} ( {\cal M})_{\sigma\lambda} \Theta^{\lambda\nu} 
   ~ p^i_{\mu} p^j_{\nu} 
\quad ,
\label{eq:inv}
\end{equation}
where we have used momentum conservation in obtaining the first equality.

The remaining terms in Eq.\ (\ref{eq:contractnp}) combine with
the restricted double sums over propagators, $G^{\prime}_P$,
for sources on the same boundary as in the derivation of the 
planar amplitude. Substituting in Eq.\ (\ref{eq:green}) 
gives the result:\footnote{Expressions for the nonplanar 
amplitude have been derived in \cite{sang,bilal,liumic} using the
background field method. A different approach, based in part on 
the algebra of open string vertex operators in a background field,
is discussed in section 2 of \cite{multi}. 
} 
\begin{eqnarray}
 {\cal A}_{NP} (1, \cdots , N) &&=
i {\tilde g}^N \delta ( \sum_{a=1}^{N} {\bf p}_a  ) 
(\alpha^{\prime})^{-(p+1)/2} {\rm det} ( {\bf 1} + {\bf B} ) 
\left [ \prod_{r=1}^{N} \int d \sigma_r \right ]
   e^{- 2 {{ {\bf k} \circ {\bar{\bf k}} }\over{l}} } 
\nonumber \\
&& \quad \times 
   \int_0^{\infty} {{dl}\over{l}} 
{{ l^{N-(p+1)/2}}\over{ \eta ( {{il}\over{2}} )^{24} }}
\prod_{i \neq j;~ i,j=1}^{N_1} \left [
e^{ {\bf k}_i \times {\bar {\bf k}}_j  \left ( \sigma_{ij} 
- \half {{\sigma_{ij}}\over{|\sigma_{ij}|}} \right ) } 
f_1(l; \sigma_{ij}) \right ]
\left [ \prod_{k=1}^{N_1}
\prod_{u=N_1+1}^{N}
f^2_2(l; \sigma_{ku}) 
\right ]
\nonumber \\
&& \quad\quad \times 
\left [ \prod_{r \neq t;~ r,t=N_1+1}^{N}
e^{ - {\bf k}_r \times {\bar {\bf k}}_t \left ( \sigma_{rt} -  
\half {{\sigma_{rt}}\over{|\sigma_{rt}|}} \right ) } 
f_1(l; \sigma_{rt}) \right ]
\quad .
\nonumber\\
\label{eq:nonp}
\end{eqnarray}
The internal momentum dependent phase factor is intriguing 
\cite{srm}, but we emphasize once again that it is dimensionless. 
Momentum transfer between boundaries is of
course best studied in the closed string channel, using 
world-sheet open-closed string duality. This will enable
us to examine carefully the limit of vanishing
momentum transfer \cite{srm}.

\subsection{Zero Momentum Transfer}
\label{sec:dual}

From the perspective of the 
closed string channel, the massless limit of the planar amplitude 
described in the previous section can be interpreted
as the limit of infinite momentum transfer: the intermediary
closed string states are infinitely massive, decoupling from the
massless fields in the open string sector. It is instructive to
consider the opposite regime of zero momentum transfer, dominated
by massless closed string exchange. We begin by performing a change
of variables, $s$$=$$2/l$, in Eq.\ (\ref{eq:nonp}), thereby 
expressing the nonplanar amplitude in a form appropriate for 
the study of zero and finite momentum transfer processes:
\begin{eqnarray}
 {\cal A}_{NP} (1, \cdots , N) &&=
i 2^{-n_p} {\tilde g}^N 
 \delta ( \sum_{a=1}^{N} {\bf p}_a  ) 
(\alpha^{\prime})^{-(p+1)/2} {\rm det} ( {\bf 1} + {\bf B} ) 
\left [ \prod_{r=1}^{N} \int d \sigma_r \right ]
   e^{- s {\bf k} \circ {\bar{\bf k}}  } 
\nonumber \\
&& \quad \times 
   \int_0^{\infty} ds 
{{ s^{-N+(p-1)/2 -12}}\over{ \eta ( is )^{24} }}
\prod_{i \neq j;~ i,j=1}^{N_1} \left [
e^{ {\bf k}_i \times {\bar {\bf k}}_j  \left ( \sigma_{ij} 
- \half {{\sigma_{ij}}\over{|\sigma_{ij}|}} \right ) } 
{\tilde f}_1( s; \sigma_{ij}) \right ]
\nonumber \\
&& \quad\quad \times 
\left [ \prod_{k=1}^{N_1}
\prod_{u=N_1+1}^{N}
{\tilde f}^2_4( s; \sigma_{ku}) 
\right ]
\left [ \prod_{r \neq t;~ r,t=N_1+1}^{N}
e^{ - {\bf k}_r \times {\bar {\bf k}}_t \left ( \sigma_{rt} -  
\half {{\sigma_{rt}}\over{|\sigma_{rt}|}} \right ) } 
{\tilde f}_1( s ; \sigma_{rt}) \right ]
\quad ,
\nonumber\\
\label{eq:nonpcl}
\end{eqnarray}
where the functions ${\tilde f}_a( s,\sigma_{ij})$ take the form:
\begin{equation}
{\tilde f}_a( s;\sigma_{ij}) \equiv 
| {{ \Theta_{a}(\sigma_{ij}, is )}\over
    {\pi s \eta^3 (is) }} |^{ 2 {\bf k}_i \cdot {\bar {\bf k}}_{j}}
\quad .
\label{eq:fmod}
\end{equation}
The Jacobi theta functions, and their modular transformations, 
can be found in \cite{polbook}. Eq.\ (\ref{eq:nonpcl}) is an 
equivalent starting point for an analysis of the nonplanar 
amplitude.

The contribution from massless closed string modes is obtained by
isolating the $O(q^0)$ term in an expansion of the integrand 
of Eq.\ (\ref{eq:nonp}) in powers of $q$$=$$e^{-2 \pi s}$. Thus, 
the field theory limit of the nonplanar amplitude with only massless 
closed string modes mediating momentum transfer takes the form:
\begin{eqnarray}
 \lim_{{\bf p}\to 0} {\cal A}_{NP} &&=
i 2^{-n_p} {\tilde g}^N 
  \delta ( \sum_{a=1}^{N} {\bf p}_a) 
(\alpha^{\prime})^{-(p+1)/2} {\rm det} ( {\bf 1} + {\bf B} ) 
\left [ \prod_{r=1}^{N} \int d \sigma_r \right ]
   \int_0^{\infty} ds s^{-N+(p-1)/2-12} 
\nonumber \\
&& \quad\quad \times 
\prod_{i \neq j;~ i,j=1}^{N_1}
\left ( 
e^{ ( {\bf k}_i \times {\bar{\bf k} }_j) ( \sigma_{ij} - \half
{{\sigma_{ij}}\over{|\sigma_{ij}|}} ) 
} 
|{{2}\over{\pi s}} {\rm Sin}(\pi\sigma_{ij})|^{ 2 
 {\bf k}_i \cdot {\bar {\bf k}}_{j} } 
\right ) 
\nonumber\\
&& \quad\quad\quad
\times ~ e^{ - s ({\bf k} \circ {\bf k}) 
}
\prod_{k=1}^{N_1}
\prod_{u=N_1+1}^{N}
\left (
 |{{1}\over{\pi s}}e^{\pi s/4} |^{ 4 {\bf k}_k \cdot {\bar {\bf k}}_{u} } 
\right )
\nonumber\\
&& \quad\quad\quad\quad
\times \prod_{r \neq t;~ r,t=N_1+1}^{N}
\left ( 
e^{ ( {\bf k}_r \times {\bar{\bf k} }_t) ( \sigma_{rt} -
\half {{\sigma_{rt}}\over{|\sigma_{rt}|}} ) 
} 
~ |{{2}\over{\pi s}} {\rm Sin}( 
\pi \sigma_{rt})|^{ 2 {\bf k}_r \cdot {\bar {\bf k}}_{t} } \right )
\nonumber\\
&& \quad\quad\quad\quad\quad
\{ 24 + 8 {\bf k}_i \cdot {\bar {\bf k}}_{j} 
 {\rm Sin}^2(\pi \sigma_{ij})  
+ 12 {\bf k}_k \cdot {\bar {\bf k}}_{u}  +  
   8 {\bf k}_r \cdot {\bar {\bf k}}_{t} {\rm Sin}^2(\pi \sigma_{rt})  
 \} 
\quad ,
\nonumber\\
\label{eq:endgarb}
\end{eqnarray}
upto terms of $O(e^{-2 \pi s})$. It is convenient to express this
in the form:
\begin{eqnarray}
 \lim_{{\bf p}\to 0} {\cal A}_{NP} &&=
i {\tilde g}^N \delta ( \sum_{a=1}^{N} {\bf p}_a) 
(\alpha^{\prime})^{-(p+1)/2} {\rm det} ( {\bf 1} + {\bf B} ) 
\left [ \prod_{r=1}^{N} \int d \sigma_r \right ]
\nonumber\\
&& \quad\quad\times
   \int_0^{\infty} ds s^{-N+(p-1)/2-12} 
e^{ - s \left [ {\bf k} \circ {\bar{\bf k}} + 
  ({\bf k} \cdot {\bar{\bf k}})\pi \right ] }
       \left ({{1}\over{s^2}}\right )^{\nu_k } 
\nonumber\\
&& \quad\quad\quad\quad\times
\left [ \prod_{i \neq j;~ i,j=1}^{N_1}
\prod_{k=1}^{N_1}
\prod_{u=N_1+1}^{N}
\prod_{r \neq t;~ r,t=N_1+1}^{N}
 F^{({\rm closed})}_0 (\sigma_a, {\bf k}_a)
\right ]
\quad ,
\label{eq:nonpfs}
\end{eqnarray}
$F^{({\rm closed})}_0$ is a polynomial function
of the ${\rm Sin}(\pi\sigma_{ab})$, $a$,$b$$=$$1$, 
$\cdots$, $N$, denoting the contribution to the amplitude from the 
massless states in the closed string sector, and 
$\nu_k$$=$$-\sum_{a=1}^N {\bf k}_a \cdot {\bar {\bf k}}_a  $.
We have changed integration variables to the closed 
string modular parameter, $s$$=$$2/l$, and the 
leading $O(e^{2\pi s})$ contribution from the closed string 
tachyon has been suppressed in writing Eq.\ (\ref{eq:nonpfs}).
The term of $O(1)$ corresponds to the exchange of massless 
closed string states.

Remarkably, the $B$ dependent terms in the exponent combine 
to give the ordinary Lorentz invariant bilinear in 
momentum transfer with Euclidean, spacelike, signature:
\begin{equation}
{\bf k} \circ {\bar{\bf k}} + \pi ({\bf k} \cdot {\bar{\bf k}}) 
   = \half \pi \alpha^{\prime} \sum_{m=0}^{(p-1)/2} 
        \left ( p_{2m} p_{2m} + p_{2m+1} p_{2m+1} \right )   
\quad ,
\label{eq:reslt}
\end{equation}
free of any dependence on the background fields! Integrating 
over the closed string modular parameter gives the result:
\begin{eqnarray}
{\cal A}_{NP}|_{{\bf p} = 0} &&=
  i {\tilde g}^N (\alpha^{\prime})^{-(p+1)/2} {\rm det} 
     ( {\bf 1} + {\bf B} ) \delta ( \sum_{a=1}^{N} {\bf p}_a ) 
       \Gamma(\nu) \left [ 
    {{2}\over{\pi\alpha^{\prime} \delta^{\mu\lambda} 
  p_{\mu} p_{\lambda}}} \right ]^{\nu}
\nonumber\\
&&\quad\quad \times
\left [ \prod_{r=1}^{N} \int d \sigma_r \right ] 
\left [ \prod_{i \neq j;~ i,j=1}^{N_1}
\prod_{k=1}^{N_1}
\prod_{u=N_1+1}^{N}
\prod_{r \neq t;~ r,t=N_1+1}^{N}
F^{({\rm closed})}_0(\sigma_a, {\bf k}_a ) \right ]
\quad ,
\label{eq:resnnp}
\end{eqnarray}
where we have defined 
$\nu$$=$${{p+1}\over{2}}$$-$$N$$-$$12$$-$$2\nu_k$. 
Note that the corrections to this result from $O(e^{-2\pi s})$ 
terms, due to the exchange of massive closed string states,
replaces the single term in the integrand of 
Eq.\ (\ref{eq:endgarb}) with a series:
\begin{equation}
\sum_{n=0}^{\infty} 
 F^{({\rm closed})}_n (\sigma_a;{\bf k}_a) 
     e^{- s[{\bf k} \circ {\bar{\bf k}} + \pi 
         ({\bf k} \cdot {\bar{\bf k}}) + 2 n\pi]}
\quad ,
\label{eq:correctnp}
\end{equation}
where the $n$th term denotes the contribution from states in the
$n$th mass level in the closed string spectrum.
Note also that Eq.\ (\ref{eq:resnnp}) holds for {\em arbitrary} 
external momenta: no kinematic constraints, or on-shell conditions, 
have been imposed on external momenta when deriving the zero momentum 
transfer limit in the closed string channel.

The amplitude has a pole whenever the momentum transfer equals the 
mass of an on-shell state in the closed string spectrum:
$\alpha^{\prime} \delta^{\mu\nu} 
p_{\mu} p_{\nu}$$=$$-4n$, with $n$ taking all integer
values including $0$.\footnote{Reference \cite{stromet} 
argues for the decoupling of the closed string sector in nonplanar
amplitudes based on the absence of poles in the expression for the 
nonplanar amplitude given in Eq.\ (2.17) of \cite{liumic}, for
S-dualized open and closed string metrics. We find no evidence for 
this in Eqs.\ (\ref{eq:resnnp}) 
and (\ref{eq:correctnp}). Notice that the open and closed string
metrics have identical spacetime dimension.} In 
particular, masses and couplings 
in the closed string sector, which appear at the loop level 
in this theory, scale naturally with respect to the bare fundamental 
string tension. In contrast, as mentioned earlier, the masses of
states in the open string sector scale naturally with the 
effective string tension. Thus, for on-shell states, respectively 
in the $n$th mass level, we have the familiar bosonic string mass 
relations:
\begin{eqnarray}
m_{{\rm closed}}^2 &&= - g^{\mu\nu} p_{\mu} p_{\nu} 
 = {{4}\over{\alpha^{\prime}}} (n -1)
\nonumber\\
m_{{\rm open}}^2 &&= - p_{\mu} \left ( {\cal M}^{-1} \right )^{\mu\nu}
  p_{\nu} = {{1}\over{\alpha^{\prime}_{{\rm eff.}} }} (n -1)
\quad ,
\label{eq:masses}
\end{eqnarray}
For convenience, we have replaced the flat spacetime metric
with its more general form, $g^{\mu\nu}$,
as explained in footnote [5]. This implies, also, the
replacement: 
${\rm det}({\bf 1}$$+$${\bf B})$$\to$${\rm det}({\bf g}$$+$${\bf B})$, 
in the definition of the effective tension 
given in Eq.\ (\ref{eq:efft}). We note that this identification 
gives a world-sheet interpretation 
of the open and closed string metrics introduced in \cite{sw}.
Note that, for a non-flat closed string metric, the closed
string masses would scale in units of a different \lq\lq
effective" string tension, namely, $\alpha^{\prime}{\rm det}(g)$. 
We return to this point in the conclusions.

We close with an important comment on the relation to the 
unoriented string.
The integral over the closed string modular parameter exhibits a
zero momentum divergence due to a tadpole when the index $\nu$ 
takes values:
\begin{equation}
\nu - 1 \equiv {{p-1}\over{2}} - 12 - N 
+ 2 \sum_{a=1}^N  {\bf k}_a \cdot {\bar {\bf k}}_{a} 
= 0 
\quad .
\label{eq:tadp}
\end{equation}
We recover the familiar tadpole both in the vacuum amplitude, 
and in any $N$ point on-shell tachyon amplitude, in the special
case of space-filling D25branes, namely, $p$$=$$25$ 
for zero momentum transfer \cite{polbook}. Note that
the tadpoles occur precisely as in the case of zero $B$ field, 
except that the coupling to the vacuum is scaled {\em up} 
in magnitude by the factor ${\rm det} ({\bf 1} $$+$$ 
{\bf B})$. 
It should be recalled that the tadpole in the vacuum amplitude for
$p$$=$$25$ can be cancelled by considering instead the unoriented 
bosonic string theory \cite{polbook}, with
$2^{13}$ D25branes and low energy gauge group $SO(2^{13})$. 
The analysis of $B$ field dependence proceeds as given
above, except that the orientation projection imposes a 
quantization condition on the values of the background $B$ field. 
The effective string tension will be raised relative to 
$(\alpha^{\prime})^{-1/2}$ as before.
Inclusion of the nonorientable diagrams does not bring 
in any new features other than are necessary to ensure 
tadpole cancellation, which works exactly as in the case
of the zero field limit. In particular, there are no additional
consistency conditions hidden in the scattering amplitudes over
what was found in the vacuum amplitude.

\subsection{Comments on Stretched Strings and the UV-IR Correspondence}
\label{sec:stre}

The momentum transfer dependent term in the nonplanar amplitude has
been given a \lq\lq stretched string" interpretation in \cite{liumic}.
We begin by pointing out that an interpretation of this term as the 
classical action for a stretched {\em
open} string fails due to a discrepancy in the dependence on the
world-sheet metric. We will use the notation of \cite{liumic}, noting 
that the fiducial cylinder metric used in \cite{liumic} coincides with 
our conventions: $g^{11}$$=$$1/l^2$,
$g^{22}$$=$$1$, with ${\sqrt {g}}$$=$$l$. The action 
for a stretched open string satisfying the boundary conditions:
\begin{equation}
X^{\mu} (\sigma^1,1) - X^{\mu} (\sigma^1,0) 
= \Theta^{\mu\nu} p_{\nu} + {\rm oscillators}, 
   \quad \quad (\Delta x)^2 = 
   - p_{\mu} \left ( \Theta G \Theta \right )^{\mu\nu} p_{\nu}
\quad ,
\label{eq:bdcls}
\end{equation}
takes the form:
\begin{equation}
S_{\rm cl.} \sim  \left (\Delta x \right )^2 l \quad ,
\label{eq:bdcl}
\end{equation}
rather than the desired term which is proportional to $1/l$! But in fact, from 
our analysis above, this term combines in the closed string channel 
with a ${\bf p}\cdot{\bf p}$ dependent term in the amplitude to give the 
ordinary 
Lorentz invariant momentum bilinear, subsequently eliminated by an 
integration over $s$$=$$1/l$. Thus, in the small momentum transfer regime, 
probed by the closed string channel, we do not find the notion of a 
stretched string interpretation appropriate.

On the other hand, in the regime of large to infinite momentum transfer, 
it may indeed be helpful to interpret the momentum transfer dependent term 
as the classical action of a stretched string as follows.
Recall that this 
regime is dominated by short open strings, and long, massive, closed strings,
with $l$$\to$$\infty$. Consider the classical solution:
\begin{eqnarray}
{\tilde x}^{\mu} (\sigma^1,\sigma^2) =&& 
     \pi \alpha^{\prime} \Theta^{\mu\nu} p_{\nu} 
          (\sigma^1 - {{1}\over{2}} ) 
\nonumber\\
\partial_1 {\tilde x}^{\mu} \partial_1 {\tilde x}_{\mu} 
=&& \pi^2 \alpha^{\prime 2} \Theta^{\mu\sigma} \left ( {\cal M} 
    \right )_{\sigma\lambda} \Theta^{\lambda \nu} p_{\mu} p_{\nu}
\nonumber\\
        =&& 2\pi\alpha^{\prime} {\bf k} \circ {\bar {\bf k}}
\quad .
\label{eq:bdclz}
\end{eqnarray}
Note that with our conventions, the constant $B$ field dependent spacetime 
metric, ${\tilde G}$, is dimensionless. Substituting in the world-sheet action, 
using the fiducial cylinder metric, gives the desired result:
\begin{equation}
S_{\rm cl.} [{\tilde z}; g] \equiv {{1}\over{4\pi\alpha^{\prime}}}
\int d^2 \sigma {\sqrt{g}} {\tilde G}_{\mu\nu} g^{ab} \partial_a {\tilde x}^{\mu}
\partial_b {\tilde x}^{\nu} = \half {{ {\bf k} \circ {\bar{\bf k}} }\over{l}}
\quad ,
\label{eq:bdcla}
\end{equation}
where we have used the fact that ${\sqrt{g}} g^{11}$$=$${{1}\over{l}}$.
The classical configuration is a \lq\lq rigid" ring-shaped annulus of 
large radius, and vanishingly small width. Large momentum transfer in the 
presence of the background $B$ field gives this annular world-sheet a 
rigidity due to the antisymmetric $B^{\mu\nu}$ coupling. We emphasize that 
while we find this helpful classical intuition, we find 
{\em no connection} 
between the stretched string and the infra-red behavior of the theory.
Nevertheless, open-closed string world-sheet duality has enabled us to 
rewrite the \lq\lq stretched string action"
--- representing a strongly UV phenomenon 
from the perspective of the noncommutative field theory limit of the open
string sector, in terms of dual closed string variables, where it 
now contributes to the poles in the nonplanar amplitude for on-shell closed 
string states--- a decidedly IR effect from the perspective of the Lorentz 
invariant, and $B$ field independent, commutative field theory of massless 
closed string states. This is simply another manifestation of the 
well-established UV-IR correspondence found in open and closed string 
perturbation theory.

We end by noting some of the unusual properties of the wavefunctions
in the presence of a constant $B$ field. Recall that the orthogonality
coefficients are not positive definite, as is apparent in Eq.\ 
(\ref{eq:params}). A consequence is an anomalously large contribution 
to the 
path integral whenever the ratio ${{n_2^2}\over{n_1^2}}$$=$$\alpha^2$.
Ordinarily, one expects the contributions from large 
eigenvalue eigenfunctions to be damped by the $\omega_{n_1n_2}^{-1}$
factor in the exponent of Eq.\ (\ref{eq:mpoint}). This is no longer true,
the exponent periodically returning to an anomalously large contribution.
In terms of the $O(e^{-\pi l})$ expansion, we note that this behavior 
represents anomalously large contributions of massive open string modes 
at specific points in the moduli space. However, these are also the 
special values at which the $(n_1,n_2)$ states become degenerate with the
$(n_1,0)$ states. So, while this semi-classical intuition is appealing, we 
should emphasize that in the computation of the Greens function we were 
able to demonstrate an almost complete cancellation of the contributions 
from these anomalous modes.

\section{Conclusions}
\label{sec:concl}

Being one-dimensional objects, strings couple naturally to an antisymmetric
two-form tensor field. The quantized particle associated with an antisymmetric
two-form potential field belongs in the same spacetime Lorentz multiplet in
string theory as the familiar graviton. In addition, superstring theories
contain additional higher p-form 
background potentials with significant consequences
for the nonperturbative dynamics of String/M theory \cite{polbook}, and, 
one would hope, equally significant consequences for string/M theoretic cosmology 
and particle physics \cite{cope,teit}. It is an important challenge to bring to 
the study of antisymmetric higher p-form gauge fields the intuition and detailed 
understanding we have of the physics of Yang-Mills gauge fields. Understanding 
the perturbative dynamics of open and closed string theory in background two-form
tensor fields is a significant step in this direction.

One of this motivations of this work was to understand more precisely
the implications of the possible noncommutativity of spacetime at
short distances for String/M theory. As mentioned at the outset, spacetimes
with constant background two-form tensor fields are but one route to
noncommutativity in String/M theory. Our work has clarified some of the
features which distinguish perturbative open and closed string theory in
a noncommutative embedding space: the finite renormalization of the bare
fundamental string tension, and the finite wavefunction renormalizations
of open string vertex operators. We have derived, in the simplified
kinematics of forward scattering with infinite momentum transfer, the precise
form of the external momentum dependent phase functions normalizing planar
amplitudes
demonstrating that they admit interpretation as a renormalization of the
wavefunctions of open string vertex operators. Using open-closed string
world-sheet duality to probe the regime of finite and zero momentum transfer,
we have shown explicitly the existence of poles in the nonplanar amplitude
when the momentum transfer equals the mass of an on-shell closed string state.
Remarkably, the field theory limit of the massless states in the closed
string sector, entering through loops of open string, is an ordinary,
Lorentz invariant, commutative quantum field theory with masses and couplings
scaling naturally in units of the fundamental string tension. This is
in sharp contrast to the field theory limit of the open string sector--- a
noncommutative field theory, with mass scale set by the effective string
tension, larger than the bare fundamental string tension, and 
vertex operator algebra and wavefunction renormalization 
derived from the star product.
We should emphasize that many of our conclusions are a 
re-interpretation in world-sheet terms of the insights in \cite{sw}.

There is a well-established UV-IR correspondence in open and closed
string theory, whereby a would-be ultraviolet phenomenon is recast as
an infrared pheneomenon in the dual string variables \cite{polbook}.
From the mass formulae for on-shell states in the open and 
closed string spectrum given in Eq.\ (\ref{eq:masses}),
it becomes clear that the notion of \lq\lq bare" and 
\lq\lq effective" string scales are, as a consequence of world-sheet 
open-closed string duality, interchangeable. While it is the 
bare fundamental string scale that appears in the world-sheet 
action, the effective string scale is actually {\em higher} in energy, 
probing shorter distance scales. In the context of Wilsonian 
renormalization in the presence of fixed background fields,
it is therefore natural to think of the, lower, fundamental closed
string scale as \lq\lq derived". Conversely, if we think in terms of
the moduli space with varying background fields, we can interpret 
the generation of an effective string scale {\em higher} than 
the closed string scale, as a {\em finite} renormalization 
effect.\footnote{Finite, only because we have, of course, assumed 
finite values of the background fields such that our world-sheet 
computations are well-founded.}

We have found no evidence in the course of our analysis to support the
notion that renormalizability in the Wilsonian sense is exotic in these 
spacetimes. The main distinction here between conclusions drawn
from noncommutative field theory\cite{srm} and from string theory lies in 
the application of open-closed string world-sheet duality. We emphasize that,
while a bilinear product identical in form to the ${\bf p} \circ {\bf p}$ 
invariant found in \cite{srm} is indeed present in the open string channel 
of the nonplanar amplitude, it is {\em dimensionless}. Using world-sheet
open-closed string duality, we have further shown that the dependence 
on internal
momentum transfer in the closed string channel is purely through the 
ordinary Lorentz invariant momentum bilinear with flat (Euclidean signature) 
metric. The nonplanar amplitude factorizes on poles at the usual on-shell
values for states in the closed string spectrum.
Thus, we find no puzzles either in the ultraviolet or the infrared
regimes. Renormalization of the open string coupling proceeds precisely as
in a commutative spacetime, taking the UV cutoff to infinity while
absorbing the divergences in the propagator in the short distance limit
into a renormalization of the bare open string coupling \cite{poltorus}. 
The low momentum transfer regime of
the nonplanar amplitudes of open string theory is instead studied by
dualizing to the closed string channel. All potential divergences appear in
the guise of an IR effect, determined by the dynamics of the zero momentum 
massless modes in the closed string sector. 
Other than the well-known divergences 
associated with the tadpole in the vacuum amplitude for the D25brane, we find
no new divergences in the closed string channel of the
nonplanar amplitude. The ${\bf p}$$\to$$0$ limit is 
benign, quite independent of the procedure by which the ultraviolet cutoff 
on the open string loop momentum is taken to infinity in 
establishing coupling constant renormalization.


\vspace{0.3in}
\noindent{\bf Acknowledgments}
We would like to acknowledge the hospitality of the CIT-USC Center
for Theoretical Physics where this work was partially completed. We 
also thank J. Michelson and the authors of \cite{multi} for their 
comments on a previous draft. This research is supported by the National 
Science Foundation grant NSF-PHY-97-22394.

\appendix
\section{\bf Computation of the Greens Function}

The infinite series appearing in the expression for the Greens function
given in Eq.\ (\ref{eq:green}) can be simplified
 by using the residue theorem and applying a Sommerfeld Watson
transform.  An elementary introduction adequate to follow the discussion can
 be found in the texts \cite{spiegelet}. The basic idea is to use the residue 
theorem to get rid of the infinite sums over the $n_1$ index.The first step
 is to rewrite the sum over $n_1$ as an integral:
\begin{equation}
\sum_{n_1=-\infty}^{\infty} f(n_1) =
\int_{C} dz ~  f(z){{{\rm Cot}(\pi z)} \over{i}} - S_1 \quad .
\label{eq:resd}
\end{equation}
In the above equation, the contour $C$ runs above the real axis, from 
$\infty+i\epsilon$ to $\infty-i\epsilon$.  The terms in the integrand are
 such that they go to zero as $z$ goes to either $+i\infty$ or $-i\infty$,
and one can perform a contour integration by closing the contour in the
upper or lower half-plane, respectively.  The part of the integral that is
evaluated by closing in the lower half-plane picks up residues from poles
at all integers $n$, which come from the factor ${\rm Cot}(\pi z)$, and 
produce the infinite sum.  In addition, there are a finite number of other
poles in the integrand whose residues contribute to the integral an extra
term $S_1$, which we subtract.  The next step is to rewrite the integrand in
such a way that any term with poles at integer $n$ goes to zero as $z$
goes to $+i\infty$, and we can close the contour in the upper half-plane.
The integral thus does not pick up any infinite sum over $n$, but rather has
a finite number of terms, which we denote by $S_2$:
\begin{equation}
\int_{C} dz ~  f(z){{{\rm Cot}(\pi z)} \over{i}} = S_2 \quad .
\label{eq:resd2}
\end{equation}
After which we can combine the two equations to get
\begin{equation}
\sum_{n_1=-\infty}^{\infty} f(n_1) = S_2 - S_1
\label{eq:sumeval}
\end{equation}
We now illustrate this process with specific examples.

To begin with, consider the series:
\begin{equation}
\sum_{n_1=-\infty}^{\infty} f(n_1) \equiv
\sum_{n_1=-\infty}^{\infty} {{e^{-2\pi i n_1 \sigma^1_-}}
	\over{n_1^2 + n_2^2l^2/4}} = 
\sum_{n_1=-\infty}^{\infty} {{{\rm Cos}(2\pi i n_1 \sigma^1_-)}
	\over{n_1^2 + n_2^2l^2/4}} \quad .
\label{eq:fsum}
\end{equation}
We can rewrite this as an integral as follows:
\begin{eqnarray}
\int_{C} dz {{{\rm Cos}(2\pi z \sigma^1_-)} \over {z^2 + n_2^2l^2/4}}
		{{{\rm Cot} (\pi z)} \over {i}} =&& 
{{1}\over {2i}}\int_{C} dz {{{\rm Cos}(2\pi z \sigma^1_-)} 
	\over {z^2 + n_2^2l^2/4}} {{e^{i\pi z} + e^{-i \pi z}} 
	\over {{\rm Sin}(\pi z)}} 
\nonumber \\
=&&  {{2 \pi i} \over {2i}} (-1) \left [ 
   {{{\rm Cos}(2\pi \sigma^1_- iln_2/2)}\over
			{2 iln_2/2}} 
{{e^{-\pi l n_2/2}}\over
			{{\rm Sin}(i \pi l n_2/2)}} 
\right ]
\nonumber \\
&&\quad 
+ {{2 \pi i} \over {2i}} \left [ 
 {{{\rm Cos}(-2\pi \sigma^1_- iln_2/2)}\over
			{-2 iln_2/2}}{{e^{-\pi l n_2/2}}\over
			{{\rm Sin}(-i \pi l n_2/2)}} \right ]
\nonumber \\
&&\quad\quad\quad 
+ {{2 \pi i} \over {2i}} \left [ {{1}\over{\pi}}
	\sum_{n_1=-\infty}^{\infty} {{{\rm Cos}(2\pi i n_1 \sigma^1_-)}
	\over{n_1^2 + n_2^2l^2/4}}  \right ] 
\quad ,
\label{eq:fsumtrans}
\end{eqnarray}
In the second line, the term in the integrand proportional to $e^{i\pi z}$ 
has been evaluated using the residue theorem by closing the contour in the
 upper half-plane, picking up a residue  at  the pole $z=in_2 l/2$ 
and an overall minus sign from clockwise integration.  
The term proportional to $e^{-i\pi z}$ has been closed in the lower-half-plane,
picking up residues at all integers $z=n_1$ as well as at $z= -i n_2l/2$.
The residues of the 
imaginary poles cancel in this expression, leaving only the sum we are 
interested in evaluating, so $S_1=0$ in this case.
 To proceed, we rearrange the integrand using
\begin{equation}
{{{\rm Cot}(\pi z)} \over {i}}= {{2e^{i\pi z}}\over{e^{i\pi z}-e^{-i \pi z}}}
	- 1  = {{e^{i\pi z}+e^{-i\pi z}}\over{2i\quad{\rm Sin}(\pi z)}}\quad ,
\label{eq:cotid}
\end{equation}
after which the integral becomes
\begin{equation}
\int_{C} dz {{{\rm Cos}(2\pi z \sigma^1_-)} \over {n_1^2 + n_2^2l^2/4}}
	\left( {{2 e^{i \pi z}} \over {e^{i \pi z}-e^{-i \pi z}}} -1 \right)
	\quad.
\label{eq:finttrans}
\end{equation}
We can now evaluate the first term by closing in the upper half-plane, where 
there is only a simple pole at $z=iln_2/2$, and we no longer have a 
contribution from the poles at $z=n$. To evaluate the second term, we need 
to split the integrand by writing 
${\rm Cos}(2\pi z \sigma^1_-) = \half e^{i 2\pi z \sigma^1_-} + \half
	e^{-i 2\pi z \sigma^1_-}$.  We then close one term in the 
upper half-plane and the other in the lower half-plane, where the terms go
to zero.  After doing this, 
one arrives at
\begin{eqnarray} 
\sum_{n_1=-\infty}^{\infty} {{e^{-2\pi i n_1 \sigma^1_-}}
	\over{n_1^2 + n_2^2l^2/4}} 
        =S_2=&& 2\pi i \left [
	-{{{\rm Cosh}(\pi n_2 l \sigma^1_-)} \over {i n_2 l}}
	{{2 e^{-\pi l n_2/2}}\over{e^{-\pi l n_2/2} - e^{\pi l n_2/2}}}
	-\half \left( -{{e^{-\pi n_2 l| \sigma^1_-|}}\over{i n_2 l}}
			+{{e^{-\pi n_2 l| \sigma^1_-|}}\over{-i n_2 l}}
	\right) \right ] \nonumber \\
	=&& {{4\pi}\over{n_2l}}\left [ {\rm Cosh}(\pi n_2l\sigma^1_-) 
	{{e^{-\pi l n_2}}\over{1 - e^{-\pi l n_2}}} 
	+ \half e^{-\pi n_2 l| \sigma^1_-|} \right ] \quad .
\label{eq:fsumtrans2}
\end{eqnarray}
Another sum we will need to evaluate is of the form 
\begin{equation}
\sum_{n_1=-\infty}^{\infty} g(n_1) \equiv 
\sum_{n_1=-\infty}^{\infty} {{e^{-2\pi i n_1 \sigma^1_- }}\over{n_2^2 - n_1^2 \alpha^2 }}
\quad .
\label{eq:gsum}
\end{equation}
We use the same trick to transform this sum, although we now have
poles along the real axis that contribute to $S_1$. 
Consider the integral
\begin{eqnarray}
{{1}\over{2i}}\int_{C} dz {{{\rm Cos}(2\pi z\sigma^1_-)}
				\over{n_2^2-z^2\alpha^2}}
{{e^{i\pi z} + e^{-i \pi z}} \over {{\rm Sin}(\pi z)}} 
=&& 
{{2\pi i}\over{2i}} 
\left [ -(0)+\sum_{n_1=-\infty}^{\infty}	
	{{{\rm Cos}(2\pi n_1 \sigma^1_-)}\over{n_2^2-n_1^2\alpha^2}}
	{{1}\over{\pi}}  
\right ]
\nonumber\\
&& \quad + {{2\pi i}\over{2i}} \left [ 
	{{{\rm Cos}(2\pi \sigma^1_- n_2/\alpha)}\over{-2 \alpha n_2}}
	{{e^{-i\pi n_2/\alpha}}\over{{\rm Sin}(\pi n_2/\alpha)}} 
\right ]
\nonumber\\
&& \quad\quad + {{2\pi i}\over{2i}} \left [ 
{{{\rm Cos}(-2\pi \sigma^1_-n_2/\alpha)}\over{2 \alpha n_2}}
	{{e^{i\pi n_2/\alpha}}\over{{\rm Sin}(-\pi n_2/\alpha)}}
	\right ]
\nonumber\\
=&&\sum_{n_1=-\infty}^{\infty}	
	{{{\rm Cos}(2\pi n_1 \sigma^1_-)}\over{n_2^2-n_1^2\alpha^2}} + S_1
\quad . 
\label{eq:gsumtrans}
\end{eqnarray}
For the first term, we've closed the contour in the upper half-plane where 
the integrand is analytic, giving a vanishing contribution to the integral. 
 The second term is closed in the lower half plane and gives contributions 
from poles at  integer values $z=n_1$ as well as $z = \pm n_2/\alpha$. 
 After rewriting the integrand using Eq.\ (\ref{eq:cotid}) we arrive at
\begin{eqnarray}
\int_{C}dz {{{\rm Cos}(2 \pi z \sigma^1_-)}\over{n_2^2-z^2\alpha^2}}
	\left({{2e^{i\pi z}}\over{e^{i\pi z}-e^{-i\pi z}}} - 1\right) =S_2
	= 2 \pi i \left [ -(0) - \half\left({{e^{-i 2\pi n_2|\sigma^1_-|/\alpha}}
	\over{-2\alpha n_2}} + {{e^{i 2\pi n_2|\sigma^1_-|/\alpha}}\over
	{2\alpha n_2}} \right) \right ] 
\quad .
\label{eq:ginttrans}
\end{eqnarray}
We then arrive at the following expression:
\begin{eqnarray}
\sum_{n_1=-\infty}^{\infty} {{e^{-2\pi i n_1 \sigma^1_- }}
\over{n_2^2 - n_1^2 \alpha^2 }} =S_2 - S_1=&&
{{\pi}\over{\alpha n_2}}\left [ {\rm Cos}(2\pi n_2 |\sigma^1_-|/\alpha)
	{\rm Cot}(\pi n_2/\alpha)+{\rm Sin}(2\pi n_2 |\sigma^1_-|/\alpha)
	\right ] \nonumber \\
&&= {{\pi}\over{\alpha n_2}}{{{\rm Cos}[(2|\sigma^1_-| - 1)\pi n_2/\alpha]}
	\over{{\rm Sin}(\pi n_2/\alpha)}} \quad .
\label{eq:gsumtrans2}
\end{eqnarray}
Using this method, one can also evaluate
\begin{eqnarray}
\sum_{n_1=-\infty}^{\infty} n_1 f(n_1) &&=
{{\sigma^1_-}\over{|\sigma^1_-|}}2\pi i
	\left({\rm Sinh}(\pi n_2l|\sigma^1_-|) 
	{{e^{-\pi l n_2}}\over{1 - e^{-\pi l n_2}}} 
	- \half e^{-\pi n_2 l| \sigma^1_-|} \right) \nonumber \\
\sum_{n_1=-\infty}^{\infty} n_1 g(n_1) &&=
 -{{\sigma^1_-}\over{|\sigma^1_-|}}
{{\pi i}\over{\alpha^2}}{{{\rm Sin}[(2|\sigma^1_-| - 1)\pi n_2/\alpha]}
	\over{{\rm Sin}(\pi n_2/\alpha)}} \quad .
\label{eq:nsums}
\end{eqnarray}

This leaves us with the sum over $n_2$$=$$0$ modes. Defining 
$\sigma \equiv \sigma^1_- + {{b}\over{l}}\sigma^2_+$, we wish to evaluate
the sum,
\begin{equation}
\sum_{n_1=-\infty }^{\infty \prime} {{e^{i \pi n_1(\alpha - 2\sigma)}}
	\over{n_1 {\rm Sin}(\pi n_1 \alpha)}} \quad .
\label{eq:zerosum}
\end{equation}
In order to exhibit the cancellation between contributions from
this term and the trigonometric terms obtained above, we resum 
using a Sommerfeld-Watson transform. We therefore consider the 
integral:
\begin{eqnarray}
I &&= {{1}\over{2i}}\int_{C}{{1}\over{z}}
	{{e^{i \pi z(\alpha - 2 \sigma)}}\over
          {{\rm Sin}(\pi z \alpha)}}
   {{e^{i \pi z}+e^{-i\pi z}}\over{{\rm Sin}(\pi z)}}
\nonumber \\
&&= {{2\pi i}\over{2i}} \left [ 
-(0) + \sum_{n_1=-\infty }^{\infty \prime}
{{e^{i \pi n_1(\alpha - 2\sigma)}}
	\over{n_1 {\rm Sin}(\pi n_1 \alpha)}} {{1}\over{\pi}} +
\sum_{n_1=-\infty }^{\infty \prime} {{e^{-i \pi n_1/\alpha(2\sigma+1)}}
	\over{n_1 {\rm Sin}(\pi n_1/ \alpha)}}{{1}\over{\pi}}
\right ]
\nonumber\\
&&\quad \quad + {{2\pi i}\over{2i}} \left [ 
{{1}\over{\alpha}}\left({{1}\over{6}} - {{\alpha^2}\over{3}} - 
	\half(2\sigma+1)^2 + \alpha(2 \sigma + 1) \right)\right ] \quad .
\label{eq:zerosumint}
\end{eqnarray}
We have closed the $e^{i \pi z}$ term in the upper half plane where
it is analytic and makes no contribution to the integral. The
$e^{-i \pi z}$ term has been closed in the lower half plane, picking up simple 
poles at $z=n$, $z=n/\alpha$ for integer $n$, and a triple pole at $z=0$.
The integrand can be rewritten as follows:
\begin{equation}
I=\int_{C}{{1}\over{z}}{{e^{i \pi z(\alpha - 2 \sigma)}}\over
{{\rm Sin}(\pi z \alpha)}}
\left( {{2 e^{i \pi z}} \over {e^{i \pi z}-e^{-i \pi z}}} -1 \right)
\quad .
\label{eq:zerosumtrans}
\end{equation}
The first term can always be closed in the upper half-plane, giving 
a zero contribution to the integral.  For the second term, we close
 in the upper half-plane for $\sigma^1_- < 0$, in the 
lower half-plane for $\sigma^1_- > 0$, giving
\begin{equation}
I = \left\{ \begin{array}{ll}
	  0\quad , & \sigma^1_- < 0   \\
	\sum_{n_1=-\infty }^{\infty \prime} {{-2i}\over{n_1}}
	e^{-2 i \pi n_1 \sigma/\alpha} + {{2 \pi}\over{\alpha}}
	(\alpha - 2 \sigma) \quad , & \sigma^1_- > 0  \end{array} 
	\right. \quad .
\label{eq:zerosumtrans2}
\end{equation}
Combining these equations leads to the expression
\begin{eqnarray}
\sum_{n_1=-\infty }^{\infty \prime}
{{e^{i \pi n_1(\alpha - 2\sigma)}}
	\over{n_1 {\rm Sin}(\pi n_1 \alpha)}}
&&= -2 \sum_{n_1=1}^{\infty \prime} {{ {\rm Cos}(\pi n_1/\alpha
	(2\sigma - {{\sigma^1_-}\over{|\sigma^1_-|}}))}\over{n_1 {\rm Sin}
	(\pi n_1 /\alpha)}} 
\nonumber\\
&&\quad - {{\pi}\over{\alpha}}
  \left({{1}\over{6}}
	 - {{\alpha^2}\over{3}} - 
	\half(2\sigma - {{\sigma^1_-}\over{|\sigma^1_-|}} )^2 
	+ \alpha(2 \sigma - {{\sigma^1_-}\over{|\sigma^1_-|}}   ) \right) 
\quad .
\label{eq:zerosumtrans3}
\end{eqnarray}
>From Eq.\ (\ref{eq:green}), the Greens function for a 
single complex scalar in constant
background $B$ field can be written as:
\begin{eqnarray}
G(\sigma^a_i,\sigma_j^a) =&& 
  {{2\alpha^{\prime}l}\over{\pi(1+b^2)}} 
\sum_{n_1=-\infty}^{\infty} e^{-2\pi i n_1 \sigma^1_-}
    \sum_{n_2=1}^{\infty} \left ( 
    {{1}\over{n_1^2 + n_2^2 l^2/4}} 
        + {{\alpha^2}\over{n_2^2 - n_1^2 \alpha^2}} 
        \right ) 
\nonumber\\
&& \quad \quad \quad \quad 
\times \left [ {\rm Cos}(\pi n_2 \sigma^2_+) - i {{\alpha n_1}\over{n_2}}
  {\rm Sin} (\pi n_2 \sigma^2_+) \right ] 
\nonumber\\
&& \quad 
      +  {{2\alpha^{\prime}l}\over{\pi}} 
           \sum_{n_1=-\infty}^{\infty}\sum_{n_2=1}^{\infty} 
             {{e^{-2\pi i n_1 \sigma^1_-}}\over{n_1^2 + n_2^2 l^2/4}} 
               {\rm Sin} (\pi n_2 \sigma^2_i) {\rm Sin} (\pi n_2 \sigma^2_j) 
\nonumber\\
&& \quad \quad \quad 
   + \sum_{n_1=-\infty }^{\infty \prime } \left [
         {{2\alpha^{\prime} b}\over{1+b^2}} e^{-2\pi i n_1 \sigma}
            {{1}\over{n_1}} {{ e^{i \pi n_1 \alpha} 
                }\over{{\rm Sin}(\pi n_1 \alpha)}} 
\right ] \quad ,
\nonumber\\
\label{eq:grexp}
\end{eqnarray}
where the prime in the last term denotes exclusion of the $n_1=0$ mode.
Rewriting the sum over $n_1$ using the Sommerfeld Watson transform gives:
\begin{eqnarray} 
G(\sigma^a_i,\sigma_j^a) &&=
 {{8 \alpha^{\prime}}\over{1+b^2}} 
\sum_{n=1}^{\infty} 
   {{1}\over{n}}  \left( {\rm Cosh} (\pi n l |\sigma^1_-|)
	{{e^{-\pi n l}}\over{1-e^{-\pi n l}}} + 
	\half e^{- \pi n l |\sigma^1_-|}\right)
      {\rm Cos} (\pi n  \sigma^2_+) 
\nonumber \\ 
&& \quad
 + {{8 \alpha^{\prime}b}\over{1+b^2}}{{\sigma^1_-}\over{|\sigma^1_-|}}
\sum_{n=1}^{\infty} {{1}\over{n}}
	\left({\rm Sinh}(\pi n l | \sigma^1_- |){{e^{-\pi n l}}
	\over{1-e^{-\pi n l}}}  - \half e^{- \pi n l|\sigma^1_-}\right)
          {\rm Sin}(\pi n \sigma^2_+ )
\nonumber\\
&& \quad \quad 
+ 8\alpha^{\prime} \sum_{n=1}^{\infty} 
  {{1}\over{n}}\left( {\rm Cosh} (\pi n l |\sigma^1_-|)
	{{e^{-\pi n l}}\over{1-e^{-\pi n l}}} + 
	\half e^{- \pi n l |\sigma^1_-|}\right) 
	{\rm Sin} (\pi n \sigma^2_i) {\rm Sin} (\pi n \sigma^2_j) 
\nonumber\\
&&\quad \quad\quad 
+ {{4 \alpha^{\prime} b}\over{1+b^2}}
  \sum_{n=1}^{\infty} {{
    {\rm Cos}((2|\sigma^1_-|-1) \pi n/\alpha) 
         }\over{n {\rm Sin}(\pi n/\alpha)}}
{\rm Cos}(\pi n \sigma^2_+)
\nonumber\\
&&\quad\quad\quad\quad 
	-{{\sigma^1_-}\over{|\sigma^1_-|}}
{{4 \alpha^{\prime} b}\over{1+b^2}}
  \sum_{n=1}^{\infty} {{
	 {\rm Sin}((2|\sigma^1_-|-1) \pi n/\alpha) 
         }\over{n {\rm Sin}(\pi n/\alpha)}}
{\rm Sin}(\pi n \sigma^2_+)
\nonumber \\
&&\quad \quad\quad \quad \quad
- {{4 \alpha^{\prime} b}\over{1+b^2}}
  \sum_{n=1}^{\infty} {{
    {\rm Cos}((2 \sigma-{{\sigma^1_-}\over{|\sigma^1_-|}}) \pi n/\alpha) 
         }\over{n {\rm Sin}(\pi n/\alpha)}}
\nonumber\\
&&\quad\quad\quad\quad\quad \quad
-{{\pi \alpha^{\prime} l}\over{1+b^2}}\left({{1}\over{6}}
	 - {{\alpha^2}\over{3}} - 
	\half(2\sigma - {{\sigma^1_-}\over{|\sigma^1_-|}} )^2 
	+ \alpha(2 \sigma - {{\sigma^1_-}\over{|\sigma^1_-|}}   ) \right)
\quad .
\label{eq:grease}
\end{eqnarray}
Notice that the last three sums, with ${\rm Sin}(\pi n/\alpha)$ in the
denominator, cancel each other out.
We can then rewrite this expression in terms of theta functions by 
using the Taylor expansions
\begin{equation}
{{e^{-\pi n l}}\over{1-e^{-\pi n l}}}= \sum_{m=1}^{\infty}e^{-\pi l n m}
\quad , \quad
\sum_{n=1}^{\infty}{{x^n}\over{n}}= - {\rm ln}(1-x) \quad ,
\label{eq:taylor}
\end{equation}
and by defining
\begin{equation}
\nu_{\pm}= {{il}\over{2}}\sigma^1_- + \half \sigma^2_{\pm} \quad ,
\tau = {{il}\over{2}} \quad .
\label{eq:def}
\end{equation}
After doing this, we arrive finally at our result:
\begin{eqnarray}
G^{\prime} (\sigma^a_i,\sigma^a_j)  &&= - 
 {{2 \alpha^{\prime} }\over{1 + b^2}} 
  {\rm ln} {{|\Theta_{1}(\nu_{+}, \tau 
     ) |^2}\over{ |\eta( \tau ) |^2 }} 
- {{2i\alpha^{\prime}b }\over{1+b^2}}
 {\rm ln} {{\Theta_{1}(\nu_{+}, \tau ) 
    }\over{ \Theta_1^{\star}(\nu_{+},\tau)}}
 - \alpha^{\prime} {\rm ln} 
 | {{\Theta_{1}(\nu_{-}, \tau ) 
   }\over{ \Theta_{1}( \nu_{+},\tau )}} |^2 
\nonumber \\ && \quad \quad 
+ {{2 \pi \alpha^{\prime} l}\over{1+b^2}}\left[\sigma^2 - {{2b\sigma}\over{l}}
 +{{2b^2}\over{3l^2}} \right ]  
\quad .
\label{eq:grbulkappy}
\end{eqnarray}

\end{document}